\documentclass[12pt]{article}
\usepackage{geometry}
\usepackage{graphicx}
\usepackage{amsmath}
\usepackage{amssymb}
\usepackage[font=small,normal, justification=raggedright,tableposition=top]{caption}
\usepackage{authblk}

\usepackage{color}

\newcommand{\pd}[2]{\frac{\partial #1}{\partial #2}} 
\newcommand{\dd}[2]{\frac{\mathrm{d} #1}{\mathrm{d} #2}} 

\newcommand{\gE}{\gamma_E}
\newcommand{\vthi}{v_{\mathrm{th}i}}

\newcommand{\rhoi}{\rho_i}

\newcommand{\kx}{k_x}
\newcommand{\ky}{k_y}
\newcommand{\kr}{k_r}
\newcommand{\kZ}{k_Z}

\newcommand{\ellx}{\ell_x}
\newcommand{\elly}{\ell_y}
\newcommand{\ellr}{\ell_r}
\newcommand{\ellZ}{\ell_Z}

\newcommand{\gEhat}{\hat{\gamma}_E}
\newcommand{\Qgb}{Q_i/Q_\mathrm{gB}}
\newcommand{\dx}{\Delta x}
\newcommand{\dy}{\Delta y}
\newcommand{\dr}{\Delta r}
\newcommand{\dZ}{\Delta Z}
\newcommand{\vz}{v_\zeta}

\newcommand{\tauc}{\tau_\mathrm{life}}

\title{Symmetry breaking in MAST plasma turbulence due to toroidal flow shear}
\author[1,2,3]{M F J Fox\thanks{michael.fox@physics.ox.ac.uk}}
\author[1,3,4]{F van Wyk}
\author[3]{A R Field}
\author[5]{Y-c Ghim}
\author[1,3]{F I Parra}
\author[1,2]{A A Schekochihin\thanks{alex.schekochihin@physics.ox.ac.uk}}
\author[3]{the MAST Team.}
\affil[1]{Rudolf Peierls Centre for Theoretical Physics, University of Oxford, Oxford OX1 3NP, UK}
\affil[2]{Merton College, Oxford OX1 4JD, UK}
\affil[3]{CCFE, Culham Science Centre, Abingdon OX14 3DB, UK}
\affil[4]{STFC Daresbury Laboratory, Daresbury WA4 4AD, UK}
\affil[5]{Department of Nuclear and Quantum Engineering, KAIST, Daejeon 305-701, Republic of Korea}
 
\renewcommand{\vec}[1]{\ensuremath{\mathbf{#1}}} 
\date{\today}

\begin{document}
\maketitle
\begin{abstract}
The flow shear associated with the differential toroidal rotation of tokamak plasmas breaks an underlying symmetry of the turbulent fluctuations imposed by the up-down symmetry of the magnetic equilibrium. Using experimental Beam-Emission-Spectroscopy (BES) measurements and gyrokinetic simulations, this symmetry breaking in ion-scale turbulence in MAST is shown to manifest itself as a tilt of the spatial correlation function and a finite skew in the distribution of the fluctuating density field. The tilt is a statistical expression of the ``shearing'' of the turbulent structures by the mean flow. The skewness of the distribution is related to the emergence of long-lived density structures in sheared, near-marginal plasma turbulence. The extent to which these effects are pronounced is argued (with the aid of the simulations) to depend on the distance from the nonlinear stability threshold. Away from the threshold, the symmetry is effectively restored. 
\end{abstract}


\section{Introduction}\label{sec:intro}
Mean flow shear associated with toroidal differential rotation is believed to play an important role in the suppression of turbulence and of the resulting transport in tokamak plasmas~\cite{Hahm1995,Burrell1997,Tynan2009}, most notably in the enhanced confinement regimes of the H-mode~\cite{Wagner1982,Wagner2007} and Internal Transport Barriers (ITBs)~\cite{Koide1994,Nazikian2005,Connor2004}. A previous study of MAST plasmas \cite{Ghim2014} found that there was a significant correlation between higher flow shear and higher temperature gradients, suggesting reduced transport; the same trend has been reported in numerical simulations \cite{Highcock2012}. However, pinning down experimentally a precise manifestation of the effect of flow shear on the local properties of the turbulence, which, presumably, causes the transport, proved to be difficult \cite{Ghim2013} and remains so. In this paper, we show, both experimentally and numerically, that the flow shear does significantly alter the properties of plasma turbulence. Besides being an essential step towards learning why and how it does so and, therefore, how turbulent transport could be manipulated in practice, it is also of fundamental physical interest to quantify and understand how the statistical properties of a turbulent plasma respond to flow shear. 


Let us start by observing that, in the absence of flow shear, the local dynamics of perturbations to an underlying Maxwellian equilibrium is constrained by a type of reflection symmetry, which follows from the up-down symmetry of the magnetic configuration \cite{Parra2011,Sugama2011}. In the presence of flow shear, this symmetry will be broken and we shall see that the signature effects of the shear can be understood in terms of this symmetry breaking. Let us explain this more quantitatively. 

Tokamak plasmas' local departures from equilibrium are described by the gyrokinetic theory (for a recent review, see \cite{Abel2013}). The gyrokinetic equation determines the evolution in time, $t$, of the perturbed distribution function of the gyrating particles, $h(x,y,z,v_\perp,v_\parallel,t)$, where $x$ is the (``radial'') coordinate perpendicular to a flux surface, $z$ is the coordinate along a field line, $y$ is the ``binormal'' (or ``poloidal'') coordinate, and ($v_\perp,v_\parallel$) are the velocity components perpendicular and parallel to the magnetic field; the third velocity-space coordinate, the gyrophase angle, is averaged over in gyrokinetics. In an up-down symmetric equilibrium, with no toroidal rotation, the gyrokinetic equation is invariant with respect to the transformations~\cite{Parra2011,Sugama2011}:
\begin{equation}
\label{symmetry-transforms}
(x,y,z) \rightarrow (-x,y,-z)\quad
v_\parallel\rightarrow-v_\parallel,\quad
h \rightarrow - h,\quad
\varphi \rightarrow -\varphi,\quad
A_\parallel \rightarrow A_\parallel,\quad
\delta B_\parallel \rightarrow -\delta B_\parallel,
\end{equation}
where $\varphi$ is the fluctuating part of the electrostatic potential, $A_\parallel$ is the perturbed parallel magnetic potential, and $\delta B_\parallel$ is the perturbed parallel magnetic field. 

\subsection{Symmetry of field distribution}
\label{sec:intro-skew}

If a stochastic (turbulent) local state of a tokamak plasma is subject to the symmetry (\ref{symmetry-transforms}), an immediate consequence is that positive and negative amplitudes of $h$, $\varphi$ and, therefore, of the density perturbation $\delta n$ (which is the most experimentally accessible fluctuating field), must occur with the same probability, i.e., the distribution of $\delta n/n$ must be symmetric (even with respect to positive and negative $\delta n/n$). A finite flow shear breaks the symmetry of the gyrokinetic equation under the radial reflection $x\rightarrow-x$, allowing this distribution to become skewed towards either over- or underdensities. 

In what follows, we will show both experimentally (Section~\ref{sec:LM-intensity-dist}) and numerically (Section~\ref{sec:gs2-skew}) that this is indeed what happens, although how pronounced the skew is depends very strongly on the distance from the nonlinear stability threshold, with symmetry essentially restored far from it (Section~\ref{sec:gs2_asymmetry}). Our experimental proxy for the density field is the fluctuating intensity field measured using the Beam Emission Spectroscopy (BES) diagnostic~\cite{Field2012} on the Mega-Amp Spherical Tokamak (MAST)~\cite{Sykes2001nf,Chapman2015}, while numerically the density field is computed using gyrokinetic simulations of a MAST plasma \cite{vanWyk2016} that employ the local flux-tube code GS2. 

\subsection{Symmetry of correlation function}
\label{sec:intro-tilt}

Another consequence of the symmetry (\ref{symmetry-transforms}), this time relating to the spatial properties of the turbulent fluctuations, is that the two-point spatial correlation function of (for example) the fluctating density field in the $(x,y)$ plane must be symmetric (even) with respect to reflection in the radial coordinate $x\rightarrow -x$. This symmetry too is broken when flow shear is present: the correlation function develops a finite tilt. 

Indeed, consider the convective time derivative of the distribution function, which incorporates toroidal rotation (in the gyrokinetic equation, this derivative is equal to the evolution operator incorporating various drifts, nonlinearities and drive terms \cite{Abel2013}): 
\begin{equation}\label{lagrangian-derivative}
\pd{h}{t} + \vec{u}\cdot\nabla h =   \pd{h}{t} + x\gE\pd{h}{y},
\end{equation}
where the toroidal rotation is locally approximated by $\vec{u}(x) = \vec{u}_0 + x \gE \hat{\vec{y}}$, we assume that we are in a frame rotating with the $x=0$ flux surface ($\vec{u}_0 = 0$), and $\gE$ is the flow shear (see the Appendix of \cite{vanWyk2016} for a concise description of this term in the local flux-tube gyrokinetic equation). Under the transformations (\ref{symmetry-transforms}), the two terms in (\ref{lagrangian-derivative}) have opposite signs, hence the breaking of the reflection symmetry. 

The effect of the shear term in (\ref{lagrangian-derivative}) on the properties of turbulence is best elicited by making the coordinate transformation into the ``shearing frame'' (e.g., \cite{Newton2010,Schekochihin2012,Fedorczak2012a})
\begin{equation}
x' = x, \quad
y' = y - x \gE t, \quad
z' = z, \quad
t' = t,
\end{equation}
under which the convective derivative (\ref{lagrangian-derivative}) becomes simply $\partial/\partial t'$. It is then legitimate\footnote{Ignoring for simplicity the effect of the magnetic shear \cite{Newton2010,Fedorczak2012a}. Note that in all measurements and simulations reported below we focus on turbulence at the mid-plane of the tokamak.}
to represent the perturbed distribution (and all other fluctuating fields) as a Fourier sum 
\begin{equation}
h = \sum_{k'} h_{k'} (t') e^{i \vec{k}'\cdot\vec{r}'} = \sum_{k'} h_{k'} (t') e^{i \vec{k}(\vec{k}',t')\cdot\vec{r}},
\end{equation}
whence, by comparing the exponents, the ``Eulerian'' wave number $\vec{k}$ can be expressed 
as a function of the ``Lagrangian'' wave number $\vec{k}'$ and time as 
\begin{equation}
\label{kx_shear}
\kx = \kx' - \ky' \gE t', \quad \ky = \ky'.
\end{equation}
Thus, in the presence of flow shear, the radial wave number grows secularly in time. 
If a typical perturbation has a life time $\tauc$, then,  using (\ref{kx_shear}), 
we can estimate the typical value of the radial wave number of turbulent fluctuations to be
\begin{equation}
\label{kx_scale}
\kx \sim \ky \gE \tauc.
\end{equation}
This tendency for perturbations with a given $\ky$ to have a certain non-zero $\kx$ 
(whose relative sign to $\kx$ is set by the sign of the shear $\gE$) manifests itself 
as a tilt in the two-point correlation function, with an angle 
\begin{equation}
\label{tilt-definition}
\Theta = -\arctan\left(\frac{\kx}{\ky}\right).
\end{equation}
The presence of a non-zero tilt $\Theta\neq0$ is a spatial manifestation of the symmetry breaking due to flow shear.

Tilting of the spatial correlation function by flow shear has been observed before, mainly in the edge of tokamak plasmas, with reports of the elongation~\cite{Alonso2006} and the breaking apart~\cite{Shesterikov2012} of turbulent perturbations. Similar observations have also been made in linear devices~\cite{Carter2009}. In the core region of a plasma, on the conventional-aspect-ratio tokamak DIII-D, a small tilt ($\sim 10^\circ$) of the spatial correlation function has been reported and attributed to flow shear~\cite{Shafer2012}. Below, we will show, again both experimentally (Section~\ref{sec:LM-spatial-corr-fun}) and numerically (Section~\ref{sec:gs2_symmetry_breaking}),  that flow shear on MAST can tilt the correlation function to large angles $>45^\circ$, making the symmetry breaking very clearly manifest and identifying the tilt as a clear signature of the presence of flow shear.  

If $\gE \tauc\ll 1$, the life time $\tauc$ will presumably be independent of $\gE$ and so  
\begin{equation}
\label{low-shear}
\Theta \sim \gE\tauc 
\end{equation}
will be proportional to the flow shear. Generally, at flow shears that are dynamically significant, $\tauc$ will have a nontrivial dependence on $\gE$. Various theories can be, and have been~\cite{Biglari1990,Shaing1990,Schekochihin2012}, constructed that provide predictions for the dependence of the life time on the flow shear. Testing such models experimentally and numerically can in principle be done by measuring the tilt $\Theta$ and then, by inverting (\ref{kx_scale}), estimating the life time of the turbulence. We will give an example of such an analysis in Sections~\ref{sec:corr_times}, \ref{sec:tilt-taulife}, and~\ref{sec:gs2_away_from_marginal}, although it will fall short of validating any specific theory due to a certain deficit of currently available experimental data. Numerically, we will discover that, like skewness, the life time of the perturbations and, therefore, the tilt angle are strong functions of the distance to the threshold, with the life time becoming shorter and the tilt gentler as this distance is increased (Section~\ref{sec:gs2_asymmetry}).\\ 

The structure of the rest of this article is as follows. In Section~\ref{sec:experiment1}, we consider three experimental cases that illustrate the effect of flow shear. In Section~\ref{sec:gksims}, the same analysis as has been performed on the experimental data in Section~\ref{sec:experiment1} is repeated on the turbulent density field from gyrokinetic simulations, the results of which are used to gain a more detailed (and more physical) understanding, through studies varying both the flow shear and the ion-temperature gradient systematically, how these two parameters affect the skewness of the distribution of $\delta n$ and the tilt of its correlation function. Finally, in Section~\ref{sec:conclusions}, we conclude with a summary of our findings, a discussion of their implications and of possible directions for future work. Supplementary technical information can be found in Appendix~\ref{sec:sources-of-skew}, where we discuss spurious sources of skewness in the experimentally measured distributions of the fluctuating intensity field, as well as the effect of the MAST BES instrument functions on this distribution. A similar, very extensive analysis regarding the tilt and other parameters of the spatial correlation function, can be found in \cite{Fox2016}. 

\section{Symmetry breaking in experimentally measured plasma turbulence in MAST}
\label{sec:experiment1}

\subsection{The MAST BES system}\label{sec:bes_intro}
The BES system on MAST~\cite{Field2012} consists of a radial-poloidal $(R, Z)$ array of $8\times4$ avalanche-photodiode (APD) detector channels~\cite{Dunai2010}, which image the South-South (SS) heating beam;\footnote{In both shots that we analyse below, the maximum SS beam power was $\simeq 2\ \mathrm{MW}$ and the maximum beam energy was $67\ \mathrm{keV}$.} the separation between the detectors is approximately $2\ \mathrm{cm}$ in either direction. The detectors register Doppler-shifted $D_\alpha$ emission, digitised at a frequency of $2\ \mathrm{MHz}$, from collisionally excited neutral-beam atoms. The fluctuating part of the intensity of this emission is proportional to the local fluctuating number density of the plasma (at fixed mean density)~\cite{Hutchinson2002}, thus enabling measurement of a turbulent field. 

The spatial localisation of the BES measurements within the beam line allows us to construct the spatial correlation function of the fluctuating density field. When this spatial correlation function is calculated (see Section~\ref{sec:BES-spatial-correlation-function}), we assume that the turbulence is homogeneous. Since there is considerable variation of some of the equilibrium quantities across the full $16\ \mathrm{cm}$ radial extent of the BES array (see Figure~\ref{fig:equilibrium_LM}), we do not use the full array, but, for the entirety of this analysis, only consider a subarray of $5\times 4$ radial-poloidal channels, which is centred at the major radii given in Table~\ref{tab:equilibria}. 

The BES diagnostic is sensitive both to the local turbulent density field and to global MHD activity~\cite{Ghim2013}. Since we are only interested in the former, we always select for analysis time intervals during which the magnetic signal, measured using a Mirnov coil at the outboard mid-plane of MAST~\cite{Hole2009}, is below a certain threshold. The time windows that we analysed for the three representative cases described in Section~\ref{sec:experimental-configuration} are: for the BLM case, $t\in 140.9210 + [0, 0.8340]\ \mathrm{ms}$; for the DLM case, $t\in368.16 + [0, 2]\ \mathrm{ms}$; for the IFS case, $t \in 125.6855 +[0, 0.5995],$ $ [1.9670, 2.7110],$ $ [2.7215,2.8095],$ $ [3.7975, 4.1255]\ \mathrm{ms}$. These time windows are long ($\gtrsim 100\times$) compared to the correlation time $\tau_c$  of the turbulence in all of these cases (see Table~\ref{tab:turb_params}). As there are also $20$ detector channels in a subarray, this means that a statistically large number of values of the turbulent density field is sampled. 

\subsection{Experimental example cases}
\label{sec:experimental-configuration}

Because the detector separation of the BES system on MAST is quite close to the typical radial correlation length of the turbulence, it is quite hard to find long intervals of BES data that would be sufficiently resolved spatially in order for reliable two-dimensional correlation functions to be obtainable (how this is done and how it is determined whether resolution limits are crossed is explained in \cite{Fox2016}). There is, therefore, relatively little data available and so it is not currently feasible to have an extensive, fully resolved parameter scan of MAST turbulence in a broad range of values of flow shear.\footnote{This is also because the range of values of shear that actually occurs in MAST is not huge anyway and, furthermore, as we will see in Section~\ref{sec:gs2_asymmetry}, the relevant parameter is likely not just $\gE$, but the distance to the nonlinear stability threshold in a multidimensional local-equilibrium-parameter space, involving at a minimum also the ion-temperature gradient.} Thus, with what we have, a practical strategy for studying the effect of flow shear on turbulence is to identify a few representative cases and compare them. Namely, we would like to examine cases where there manifestly is or is not flow shear present and ask whether they are different in a clearly measurable and qualitatively understandable way.

\begin{figure}
\centerline{\includegraphics[width=0.8\textwidth]{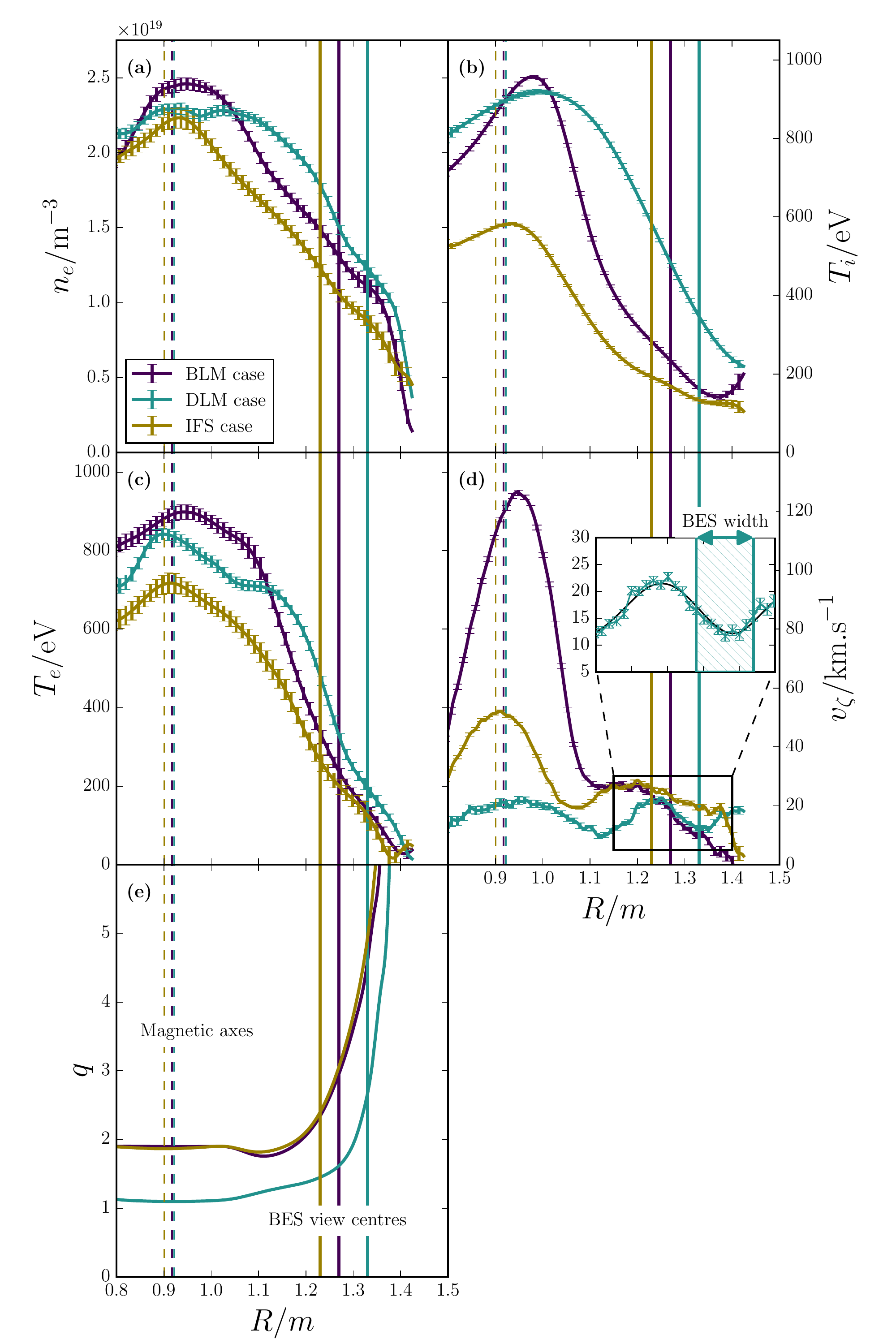}}
\caption{Cubic-spline-fitted equilibrium profiles for the BLM, DLM and IFS cases (described in Section~\ref{sec:experimental-configuration}): (a) electron density, measured using Thomson scattering (TS)~\cite{Scannell2008,Scannell2010}, (b) ion temperature, measured using Charge-Exchange Recombination Spectroscopy (CXRS)~\cite{Conway2006}, (c) electron temperature (from TS), (d) toroidal velocity (from CXRS), and (e) $q$-profile, reconstructed by EFIT constrained by the motional-Stark-effect (MSE)~\cite{Conway2010} measurements. The vertical solid lines indicate, for each of the three cases, the centre of the BES viewing location for the subarray being used (the width of the subarray is $10~\mathrm{cm}$). The vertical dashed lines indicate the position of the magnetic axis in each case. The black curve in the inset to (d) is the cubic-spline fit used for the calculation of (\ref{gammaE_def}). \label{fig:equilibrium_LM}}
\end{figure}

\begin{table} 
	\centering

	\begin{tabular}{lrrrrr} 
		Quantity&Symbol&BLM&DLM&IFS&EGK \\ \hline 
		Shot number&Shot& $28155$ & $28155$ & $27278$ & $27268$   \\ 
		Time into shot$/\mathrm{s}$& & $ 0.141$ & $ 0.369$ & $ 0.127$ & $ 0.250$    \\ 
		Major radius at centre of BES view$/\mathrm{m}$&$R$& $ 1.27$ & $ 1.33$ & $ 1.23$ & $ 1.32$  \\ \hline 
		Normalised flow shear (\ref{gammaE_def})&$\hat{\gamma}_E$& $ 0.054$ & $ 0.022$ & $ 0.034$ & $ 0.160$ \\
                Electron-temperature gradient&$a/L_{T_e}$& $ 4.8$ & $ 5.3$ & $ 4.3$ & $ 5.8$   \\ 
                Ion-temperature gradient&$a/L_{T_i}$& $ 3.1$ & $ 3.3$ & $ 1.9$ & $ 5.1$   \\ 
                Electron-density gradient&$a/L_{n_e}$& $ 1.7$ & $ 1.9$ & $ 2.1$ & $ 2.6$   \\  
		Electron-ion temperature ratio&$T_e/T_i$& $ 1.0$ & $ 0.56$ & $ 1.4$ & $ 1.1$   \\ 
		$\beta=8\pi n_i T_i/B^2$&$\beta$& $ 0.010$ & $ 0.008$ & $ 0.010$ & $ 0.005$   \\ 
		Minor radius/minor radius of LCFS&$r/a$& $ 0.63$ & $ 0.71$ & $ 0.55$ & $ 0.80$   \\ 
		Safety factor&$q$& $ 3.0$ & $ 3.0$ & $ 2.5$ & $ 2.3$   \\ 
		Magnetic shear&$\hat{s}$& $ 2.2$ & $ 5.2$ & $ 1.7$ & $ 4.0$   \\ 
		Ion-ion collisionality&$\nu_{ii*}$& $ 0.014$ & $ 0.006$ & $ 0.020$ & $ 0.020$   \\ 
		Electron-ion collisionality&$\nu_{ei*}$& $ 0.79$ & $ 0.85$ & $ 0.74$ & $ 0.59$   \\ 
		Elongation&$\kappa$& $ 1.6$ & $ 1.4$ & $ 1.4$ & $ 1.5$   \\ 
		Derivative of elongation&$\kappa'$& $ 0.55$ & $ 0.73$ & $ 0.24$ & $ 0.45$   \\ 
		Triangularity&$\delta$& $ 0.22$ & $ 0.19$ & $ 0.21$ & $ 0.21$   \\ 
		Derivative of triangularity&$\delta'$& $ 0.081$ & $ 0.86$ & $ 0.49$ & $ 0.46$   \\ 
		Shafranov shift&$\Delta'$& $ -0.33$ & $ -0.48$ & $ -0.27$ & $ -0.31$   \\ 
		Mach number&$M$& $ 0.12$ & $ 0.074$ & $ 0.19$ & $ 0.43$  \\ \hline 
		Ion gyroradius$ / \mathrm{cm}$&$\rho_i$& $ 0.88$ & $ 1.1$ & $ 0.78$ & $ 0.66$   \\ 
		Time normalisation $ / \mu\mathrm{s}$&$a/\vthi$& $ 3.8$ & $ 3.2$ & $ 4.4$ & $4.0$   \\ 
		Pitch angle of magnetic field$/\mathrm{deg}$&$\alpha$& $ 23$ & $ 30$ & $ 21$ & $ 34$   \\ 
		Ion temperature$/\mathrm{eV}$&$T_i$& $ 230$ & $ 350$ & $ 190$ & $ 220$   \\ 
		Electron temperature$/\mathrm{eV}$&$T_e$& $ 250$ & $ 200$ & $ 270$ & $ 240$   \\ 
		Electron density$/10^{19}\ \mathrm{m}^{-3}$&$n_e$& $ 1.3$ & $ 1.2$ & $ 1.2$ & $ 1.2$   \\ 
		Toroidal velocity (from CXRS)$/\mathrm{km/s}$&$v_\zeta$& $ 18$ & $ 14$ & $ 26$ & $ 62$   \\ 
		Toroidal velocity (from BES)$/\mathrm{km/s}$&$v_\zeta^\mathrm{BES}$& $13\pm9$ & $8\pm4$ & $ 19\pm4$ & $-$   \\  \\ 
	\end{tabular} 
	\caption{Local equilibrium parameters for each of the experimental cases analysed (Section~\ref{sec:experimental-configuration}). The BLM and DLM cases are the main comparison cases discussed in Sections~\ref{sec:LM-spatial-corr-fun} and~\ref{sec:LM-intensity-dist}; the IFS case is first introduced in Section~\ref{sec:IFStilt}. The EGK case gives the parameters for the gyrokinetic simulations analysed in Section~\ref{sec:gksims} and is also discussed there. Profiles of $n_e$, $T_i$, $T_e$, $\vz$, and $q$ are plotted in Figure~\ref{fig:equilibrium_LM} for the BLM, DLM and IFS cases. The numerical values for these three cases are average quantities over the width of the BES subarray, whilst the values for the EGK case are taken at the centre of the BES array (those were the values used in simulations). \label{tab:equilibria}} 
\end{table} 

The answer is that they are and to demonstrate this, we compare measurements made using the BES system on MAST from two times in the shot \#28155. This shot is particularly useful for distilling the effect of flow shear because at $t=0.36~\mathrm{s}$ a locked mode occurs, braking the rotation of the plasma~\cite{Howell2007}. The two times that we consider are before the locked mode (BLM) at $t=0.141~\mathrm{s}$ and during the locked mode (DLM) at $t=0.369~\mathrm{s}$. Measurements of the toroidal velocity using the charge-exchange recombination spectroscopy (CXRS) diagnostic are plotted in Figure~\ref{fig:equilibrium_LM}(d) for these two times. Before the locked mode, an ITB is present at the major radius $R\simeq1.1\ \mathrm{m}$, resulting in a steep gradient in the toroidal velocity, however, the BES measurement location in that experiment is outside of the ITB at $R=1.27\ \mathrm{m}$, where the flow shear is less strong. During the locked mode, the ITB no longer exists and the toroidal rotation profile flattens. The BES view is centred at $R=1.33\ \mathrm{m}$ for this latter analysis. In the centre of the BES viewing location, a magnetic island forms, within which there is no gradient in the toroidal rotation: see the inset of Figure~\ref{fig:equilibrium_LM}(d). Therefore, by studying this shot, we have access to two clearly different regimes: one with (BLM) and one without (DLM) the flow shear. 

The BLM case is quite typical of the sheared cases that can be found in the BES measurements data base---and most of this data base consists of sheared cases. Indeed, the BES diagnostic requires the SS neutral beam in order to measure the turbulence in MAST and the tangentially directed beam provides a torque to the plasma, causing it to rotate. Therefore, there is almost always some level of flow shear acting on the turbulent fluctuations measured by the BES. The DLM case is, thus, rather special, but that is the point: turbulence in MAST is rarely without a flow shear and so any such situation provides a very useful opportunity for a clean comparison. A third experimental example, which will be introduced in Section~\ref{sec:IFStilt}, is an intermediate-flow-shear (IFS) case, taken from shot \#27278 at $t=0.128\ \mathrm{s}$, with the BES measurement taken at $R=1.23\ \mathrm{m}$. It represents a less clear-cut situation, which we include to highlight the range of behaviour that one encounters in MAST. While we do not claim to have identified all of the interesting cases, we did analyse a large number of intervals and the examples we have chosen do appear to us to be a worthwhile selection.  

The profiles of the electron number density $n_e$, ion ($T_i$) and electron ($T_e$) temperatures and the safety factor $q$, in addition to the toroidal velocity $\vz$ already discussed above, are plotted in Figure~\ref{fig:equilibrium_LM} for all three cases. In Table~\ref{tab:equilibria}, a full set of equilibrium parameters is provided: the values quoted there are averages over the width of the BES subarray that we use for measurements (see Section~\ref{sec:bes_intro}). All three cases are plasmas with double-null-divertor (DND) geometries, and, therefore, the equilibria must be up-down symmetric, making the symmetry considerations of Section~\ref{sec:intro} relevant.  

Together, the three cases described above provide examples across a typical (for MAST) range of values of the normalised flow shear
\begin{equation}
\label{gammaE_def}
\gEhat = \frac{\gE}{\vthi/a} = -\frac{r}{q} \dd{\omega}{r}\frac{a}{\vthi},
\end{equation}
where $q$ is the safety factor, $r$ is the distance from the magnetic axis to the BES viewing location, $\vthi = \sqrt{2 T_i/m_i}$ is the ion thermal speed, $T_i$ the ion temperature, $m_i$ the Deuteron mass, $a$ the minor radius of the last-closed flux surface (LCFS), $\omega=\vz/R$ the toroidal rotation frequency, $\vz$ the toroidal velocity and $R$ the major radius at the BES viewing location (these quantities can be found in Table~\ref{tab:equilibria}). The definition (\ref{gammaE_def}) is the same as that used as an input parameter for the gyrokinetic simulations using GS2, however, it is different from the physical flow shear by a factor of $qR B_{\rm p}/r B$, where $B$ is the total magnitude of the magnetic field, and $B_{\rm p}$ its poloidal component. In the large-aspect-ratio limit, this factor is close to unity, but, for a spherical tokamak such as MAST, there is a significant difference. For consistency with the standard definition of $\gE$ used in simulations \cite{vanWyk2016} (Section~\ref{sec:gksims}), we will always use (\ref{gammaE_def}) as our flow-shear parameter. 

Note that, from inspecting the raw velocity profile in Figure~\ref{fig:compare_LM_shots}(d), we know that the flow shear in the DLM case is approximately zero across half of the BES subarray, however, the value for the DLM flow shear given in Table~\ref{tab:equilibria} is nearly two thirds of the flow shear for the IFS case. This apparent discrepancy occurs because a cubic-spline fit (see inset of Figure~\ref{fig:equilibrium_LM}(d)) is performed in order to calculate the radial derivative of the toroidal velocity, and the spline fit underestimates the flatness (or, equivalently, the width) of the region due to the magnetic island. The resulting value of the flow shear, calculated as the average over the BES subarray, depends sensitively on the point at which the spline fit determines the minimum to occur, as the sign of the flow shear changes at this point, resulting in significant uncertainty in the calculated value of the flow shear in this case.

\subsection{Two-dimensional spatial structure of the turbulence}
\label{sec:LM-spatial-corr-fun}

\subsubsection{Instantaneous turbulent density field}
\label{sec:bes_2d_frames}

\begin{figure}
	\centering
	\includegraphics[width=\linewidth]{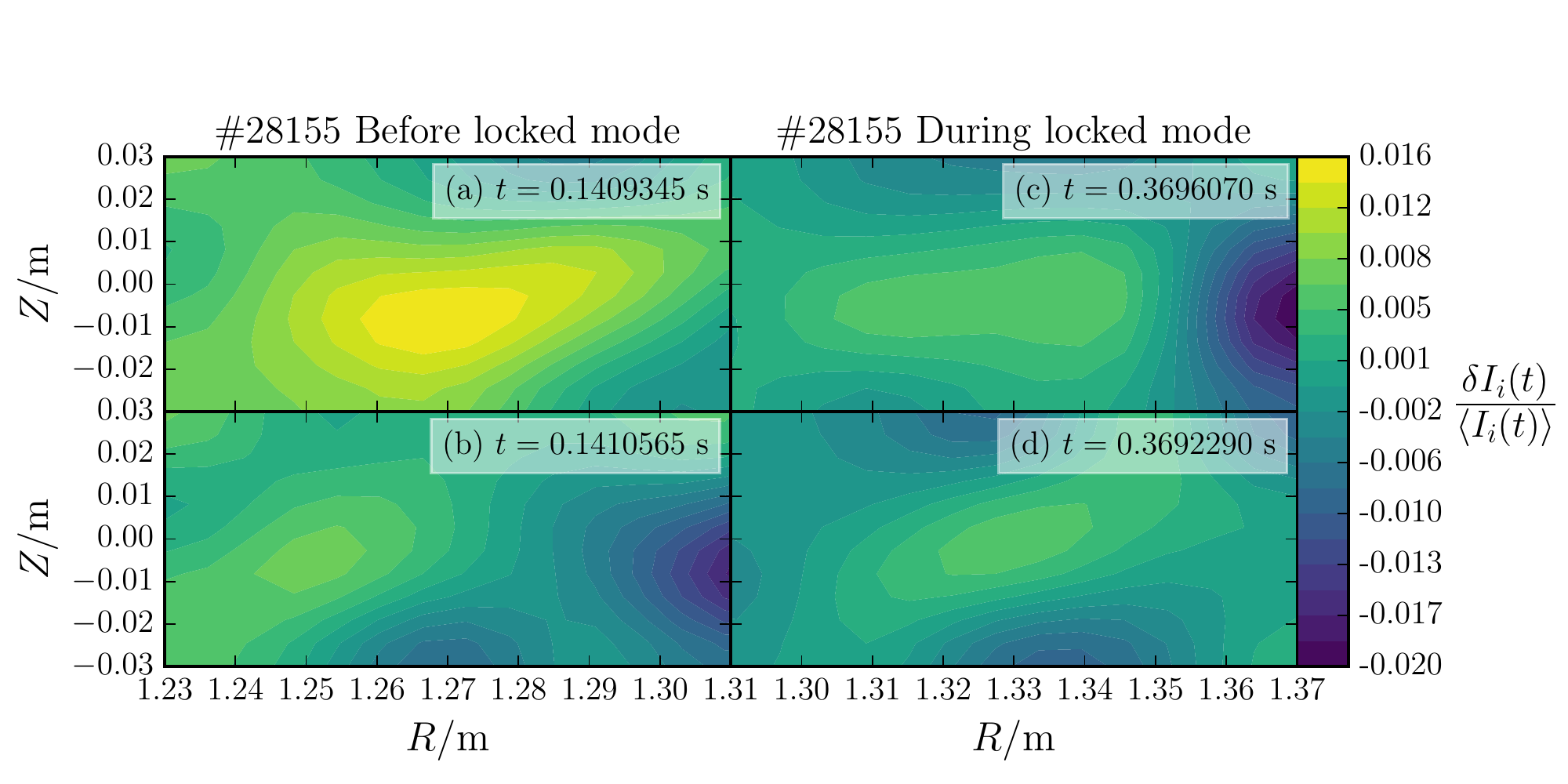}
	\caption{Snapshots of the raw (but 2D-spline-interpolated) BES fluctuating-intensity signal $\delta I_i/\langle I_i\rangle$ from shot \#28155 at: (a) $t=0.1409345\ \mathrm{s}$, (b) $t=0.1410565\ \mathrm{s}$, (c) $t=0.3696070\ \mathrm{s}$, (d) $t=0.3692290\ \mathrm{s}$. Times (a) and (b) occur during the BLM time period (significant flow shear) and times (c) and (d) occur during the DLM time period (no flow shear). The correlation functions of these two cases are given in Figure~\ref{fig:compare_LM_shots}, which include these snapshots in the temporal average. \label{fig:raw_data}}
\end{figure}

Before discussing statistical properties of turbulence, let us inspect the instantaneous spatial structure of the fluctuating density field during the time windows being studied. In Figure~\ref{fig:raw_data}, example snapshots are plotted of the BES-measured intensity amplitude at certain chosen times. Each channel has been normalised by its own mean, so these are essentially snapshots of the relative density perturbation $\delta n/n$. 

While we know (and it will also become very obvious in Section~\ref{sec:BES-spatial-correlation-function}) that the BLM case [Figures~\ref{fig:raw_data}(a) and (b)] has flow shear and the DLM case [Figures~\ref{fig:raw_data}(c) and (d)] does not, it is not easy (indeed, well-nigh impossible) to deduce that by inspection of such snapshots or by watching the full time-evolution sequences for these two cases. Indeed, comparing Figures~\ref{fig:raw_data}(a) and (c), we see that similar untilted (and so seemingly unsheared) structures occur in both the BLM and the DLM cases; conversely, comparing Figures~\ref{fig:raw_data}(b) and (d), we also see that similar tilted structures also occur in both cases.\footnote{In the case with no flow shear, instantaneous tilted structures could be due to (and indeed be a signature of) a fluctuating zonal flow shearing the density field; this also happens in numerical simulations with $\gE=0$, as discussed in Section~\ref{sec:ZF}.} Therefore, the morphology of particular realisations of the turbulence does not reveal in any especially glaring way that turbulence is different between the two cases. Statistically, however, it is very different, as we are about to discover. 

\subsubsection{Spatial correlation function}
\label{sec:BES-spatial-correlation-function}

We now characterise the average spatial structure of turbulence using the spatial correlation function of the BES-measured intensity signal. Denoting by $I_i(t)$ the time-dependent intensity measured by the BES detector channel~$i$, we split this signal into mean (averaged over time) and fluctuating parts, 
\begin{equation}
\label{Isplit}
I_i(t) = \langle I_i\rangle + \delta I_i(t). 
\end{equation}
The two-point correlation matrix of the fluctuating intensity field is then 
\begin{equation}
\label{correlation_def}
C_{ij} =\frac{\langle \delta I_i(r_i, Z_i) \delta I_j(r_j, Z_j) \rangle}{\sqrt{\langle \delta I_i^2(r_i, Z_i)\rangle \langle \delta I_j^2(r_j, Z_j) \rangle}},
\end{equation}
where $r_i$, and $Z_i$ are the radial and poloidal (vertical) locations of the detector channel~$i$, respectively, and the angle brackets again denote time averages (the auto-correlation functions for each detector, calculated for $i=j$, are corrected for photon-noise effects as described in~\cite{Ghim2013} and at the beginning of Section~\ref{sec:LM-intensity-dist}). Assuming the turbulence is (approximately) spatially homogeneous across the BES subarray used for our analysis (see Section~\ref{sec:bes_intro}), its two-point correlation function depends only on the relative distances between channels $\dr= r_i-r_j$ and $\dZ = Z_i-Z_j$. We reconstruct this correlation function from the correlation matrix (\ref{correlation_def}) by considering the relevant ranges of values of $\dr$ and $\dZ$ and using the following fitting function 
\begin{equation}
\label{fit_fun}
C(\dr, \dZ) = p + (1-p)\exp \left[ -\left( \frac{\dr^2}{\ellr^2} + \frac{\dZ^2}{\ellZ^2} \right)\right]\cos\left[ \kr \dr+ \kZ \dZ \right].
\end{equation}
The parameter $p$ is used to account for global modes~\cite{Ghim2013}, as well as for fluctuations in the neutral beam~\cite{Moulton2015}. As we have selected times during which there is no MHD activity (see Section~\ref{sec:bes_intro}), the value of $p$ should be small. Indeed, in all measured cases it is at most a few percent, therefore also indicating that neutral-beam fluctuations do not affect the measurements. 

The fit (\ref{fit_fun}) allows us to extract the spatial correlation parameters of the turbulence: the radial, $\ellr$, and poloidal, $\ellZ$, correlation lengths and the radial, $\kr$, and poloidal, $\kZ$, wave numbers. Operationally, in order to constrain the fit (\ref{fit_fun}), we fix the product $\kZ\ellZ$, which is determined from the two-time, single-point correlation function of the intensity field (see Section 2.4 and Appendix A of \cite{Fox2016}). To obtain correct values of the spatial correlation parameters of the true density field, one must take account of the finite spatial resolution of the BES system, which is quantified in terms of the point-spread functions (PSFs) of its detector channels. How to do this systematically is worked out in \cite{Fox2016}; here $\ellr$, $\ellZ$, $\kr$, and $\kZ$ are all corrected for PSF effects using this technique. 

\begin{figure}
\centerline{\includegraphics[width=\textwidth]{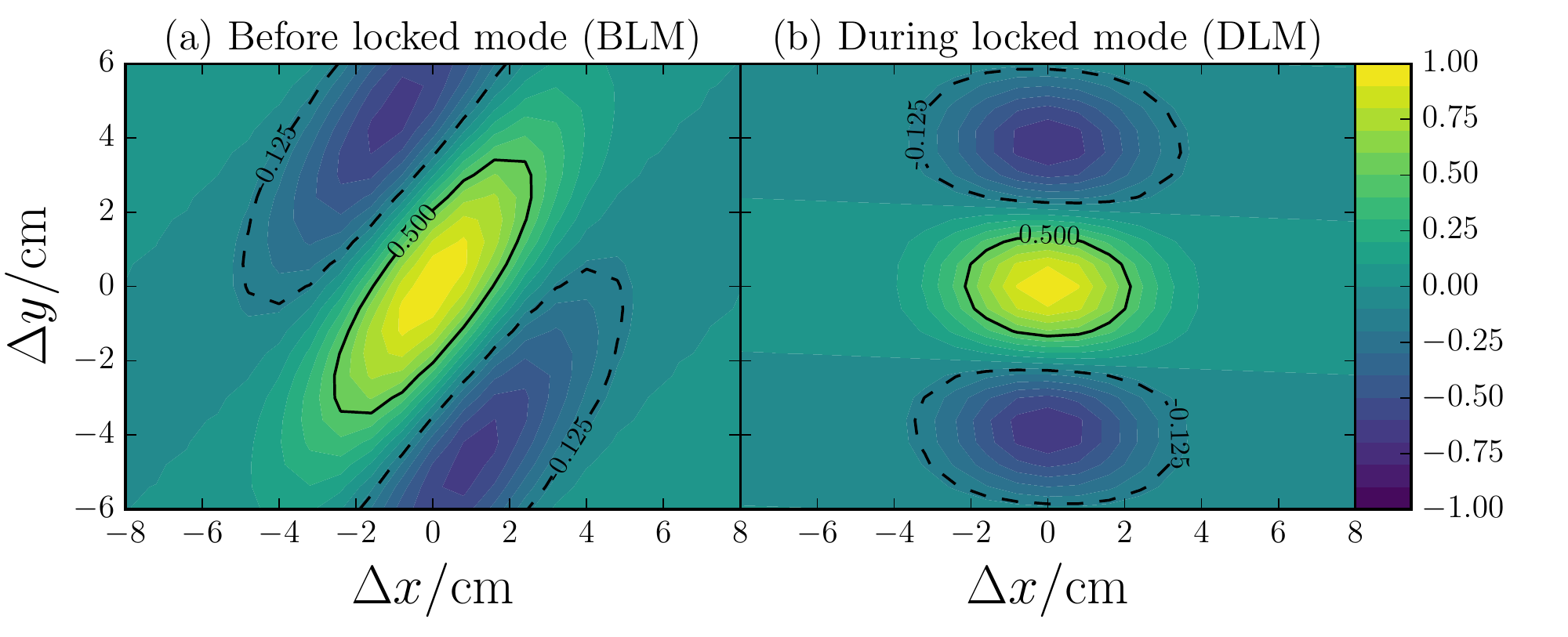}}
\caption{Spatial two-point correlation function (\ref{plasma-corr-fun}) of the fluctuating density field for (a) the case before the locked mode (BLM, with flow shear) and (b) the case during the locked mode (DLM, no flow shear), both described in Section~\ref{sec:experimental-configuration}. Table~\ref{tab:turb_params} shows the parameters of these correlation functions calculated by fitting (\ref{fit_fun}) to the spatial correlation function (\ref{correlation_def}), correcting for PSF effects, and transforming into the $(x,y)$ coordinates perpendicular to the magnetic field. These correlation functions should be compared to the correlation functions of the numerically simulated turbulence in Figure~\ref{fig:gs2_compare_correlation_functions}. \label{fig:compare_LM_shots} }
\end{figure}

Physically, we would like to work with the correlation function in terms of point separations in the plane $(x,y)$ perpendicular to the mean magnetic field, where $x$ is the coordinate perpendicular to the flux surface (so, in the outboard midplane of the tokamak, the same as the radial coordinate) and $y$ is the ``binormal'' coordinate, perpendicular both to the field and to the $x$ direction. Assuming that the correlation length of the fluctuating density field along the magnetic field is long compared to $\ellZ$, we then convert the correlation function (\ref{fit_fun}) to  
\begin{equation}
\label{plasma-corr-fun}
C(\dx, \dy) = \exp \left[ -\left( \frac{\dx^2}{\ellx^2} + \frac{\dy^2}{\elly^2} \right)\right]\cos\left[ \kx \dx+ \ky \dy \right],
\end{equation}
where $\ellx = \ellr$, $\elly = \ellZ\cos\alpha$, $\kx = \kr$, $\ky = \kZ/\cos\alpha$, and $\alpha = \arcsin(B_{\rm p}/B)$. These correlation lengths and wave numbers are given in Table~\ref{tab:turb_params}. The spatial correlation function (\ref{plasma-corr-fun}) is plotted in Figures~\ref{fig:compare_LM_shots}(a) and (b) for the BLM and DLM cases, respectively. 

\begin{table} 
	\centering
	\begin{tabular}{r|ll|l} 
		Name    & BLM & DLM&    IFS  \\ 
		$\gEhat$ & $ 0.054\pm0.006$ & $ 0.02\pm0.04$ & $ 0.034\pm0.012$  \\ 
                \hline\rule{0pt}{1.2em}
                $\overline{\delta n/n}$ & $0.011 \pm 0.002$  & $0.03 \pm 0.01$ &  $0.020 \pm 0.003$\\
		Skewness  & $ 0.21\pm0.03 $& $ 0.01\pm0.03$ &$ -0.17\pm0.02$\\
                \hline\rule{0pt}{1.2em}
		$\ellx/\rho_i$    & $3.9 \pm 1.5$ & $ 2.4 \pm 0.6$ & $ 3.3 \pm 0.2$    \\ 
		$\elly/\rho_i$ &$9.7 \pm 0.8$  & $ 6.2 \pm 0.7$ &  $ 10.7 \pm 1.2$   \\ 
		$\kx\rho_i$    & $-0.51 \pm 0.14$  & $0.03 \pm 0.20$ &$ -0.21 \pm 0.08$   \\ 
		$\ky\rho_i$  &  $ 0.43 \pm 0.03$& $ 0.82 \pm 0.03$ &  $ 0.43 \pm 0.05$ \\ 
		Tilt: $\Theta$/deg    & $50\pm8$ & $-2\pm15$&$26\pm9$ \\
                \hline\rule{0pt}{1.2em}
		$\tauc/(a/\vthi)$ & $22 \pm 7 $ & $-$ & $14\pm8$ \\
		$\tau_\mathrm{c}/(a/\vthi)$ & $3\pm2$  & $6\pm2$& $3\pm1$ \\
		$\tau_\mathrm{c}/\mu\mathrm{s}$ & $12\pm9$  & $20\pm6$& $19\pm4$ \\
	\end{tabular} 
	\caption{Correlation parameters for BLM, DLM, and IFS cases, all described in Section~\ref{sec:experimental-configuration}. The correlation functions (\ref{plasma-corr-fun}) with these parameters are shown in Figures~\ref{fig:compare_LM_shots} and~\ref{fig:IFS_shot}. All lengths are normalised by the ion gyroradius $\rhoi = \vthi/\Omega_i$ ($\vthi$ is the ion thermal speed, $\Omega_i$ the ion cyclotron frequency), all times to $a/\vthi$; the values of both of these can be found in Table~\ref{tab:equilibria}. The rms amplitude $\overline{\delta n/n}$ of the fluctuating density field is calculated from the rms amplitude $\overline{\delta I/I}$ of the fluctuating intensity field, as defined by (\ref{std-dev}), using the method described in Section 7.1 of~\cite{Fox2016}. The skewness is calculated using (\ref{skew-def}) for the distribution of $\delta I$. The life time $\tauc$ is defined in (\ref{tauc-def}); the correlation time $\tau_\mathrm{c}$ was calculated using the CCTD method~\cite{Durst1992,Ghim2012PPCF}. \label{tab:turb_params}} 
\end{table}

\subsubsection{Symmetry breaking}
\label{sec:sb-tilt}

The difference between the BLM case (with flow shear) and the DLM case (without flow shear) is manifest: the BLM correlation function is highly tilted whereas the DLM one is very nearly even in $\dx$. The tilt can be defined in terms of the tilt angle (\ref{tilt-definition}), where we now use the $\kx$ and $\ky$ parameters of the fitting function (\ref{plasma-corr-fun}). The values of $\Theta$ are given in Table~\ref{tab:turb_params}. Thus, in line with the theoretical discussion in Section~\ref{sec:intro-tilt}, the statistical symmetry of the plasma turbulence under radial reflection is broken in the presence of flow shear. 

\subsubsection{Correlation times and flow shear}
\label{sec:corr_times}

As promised in Section~\ref{sec:intro}, we may calculate the effective ``life time'' of a sheared structure, defined, in view of (\ref{kx_scale}), by
\begin{equation}
\label{tauc-def}
\tauc \equiv \left|\frac{\kx}{\gE\ky}\right| = \left|\frac{\tan\Theta}{\gE}\right|. 
\end{equation}
The values of this quantity for our experimental cases are given in Table~\ref{tab:turb_params}. We also give there the correlation time $\tau_\mathrm{c}$ of the fluctuating density field determined from its time-correlation function using the cross-correlation time delay (CCTD) method~\cite{Durst1992,Ghim2012PPCF}. We see that the values of $\tauc$ and $\tau_\mathrm{c}$, while not wildly disparate, cannot really be declared to agree. We are going to be relaxed about this problem: first, because both the CXRS value of $\gE$ and the CCTD value of $\tau_\mathrm{c}$ are subject to large errors (regarding the reliability of the CCTD method in the presence of non-negligible PSFs, see Section~9 of \cite{Fox2016}), and secondly, because (\ref{kx_scale}) is obviously a relation allowing an arbitrary constant of order unity. However, if the measurement of $\gE$ is viewed as reliable, the life time defined by (\ref{tauc-def}) can be meaningfully used as an effective measure of the (Lagrangian\footnote{The correlation time appearing in (\ref{kx_scale}) is probably best interpreted as a Lagrangian correlation time (life time of a moving and sheared ``eddy'', which is reflected in our notation for $\tauc$), whereas the correlation time $\tau_\mathrm{c}$ determined from the BES-measured time correlation function is obviously the Eulerian correlation time (albeit in the frame moving with the overall mean toroidal rotation velocity \cite{Ghim2012PPCF}). While in conventional (Kolmogorov-like) strong turbulence, one expects $\tauc\sim \tau_\mathrm{c}$, the relationship between the two times may be more subtle in a turbulent state where long-lived coherent structures play a prominent role---it was argued in \cite{vanWyk2016} that MAST turbulence is in just such a state close to the (nonlinear) stability threshold (this will be further discussed in Section~\ref{sec:physics-skew}).}) correlation time for comparative studies of turbulence at different equilibrium-parameter values, as it will be in Sections~\ref{sec:tilt-taulife} and \ref{sec:gs2_away_from_marginal}.

Correlation functions of MAST turbulence generically exhibit a tilt when flow shear is present, with a range of values of $\Theta$---we have checked this in a number of experimental cases, not shown here. This is a reliable feature, to the extent that it could be thought of as a way to find out whether a flow shear is present.\footnote{Note that the flow shear need not be dynamically significant in order for the correlation function to be tilted. At low shear, the tilt (small in this limit) would be a ``passive'' feature, proportional to $\gE$, as per (\ref{low-shear}); see further discussion in Section~\ref{sec:gs2_away_from_marginal}.} Given that the values of $\gE$ might not be computed very precisely from CXRS measurements of the rotation profiles, one might wonder whether in fact a more reliable (and dynamically relevant) measure of the local flow shear would be obtained from the tilt angle of the spatial correlation function and the correlation time $\tau_\mathrm{c}$ computed from the time-correlation function, viz., $\gE^\mathrm{(eff)} = {\tan\Theta}/{\tau_\mathrm{c}}$ (ignoring here the difference between Lagrangian and Eulerian correlation times; note that $\gE^\mathrm{(eff)}$ could also be the local shear associated with zonal flows, as will be discussed in Section~\ref{sec:ZF}).
 
\subsubsection{Case of intermediate flow shear} 
\label{sec:IFStilt}

\begin{figure}
\centerline{\includegraphics[width=0.6\textwidth]{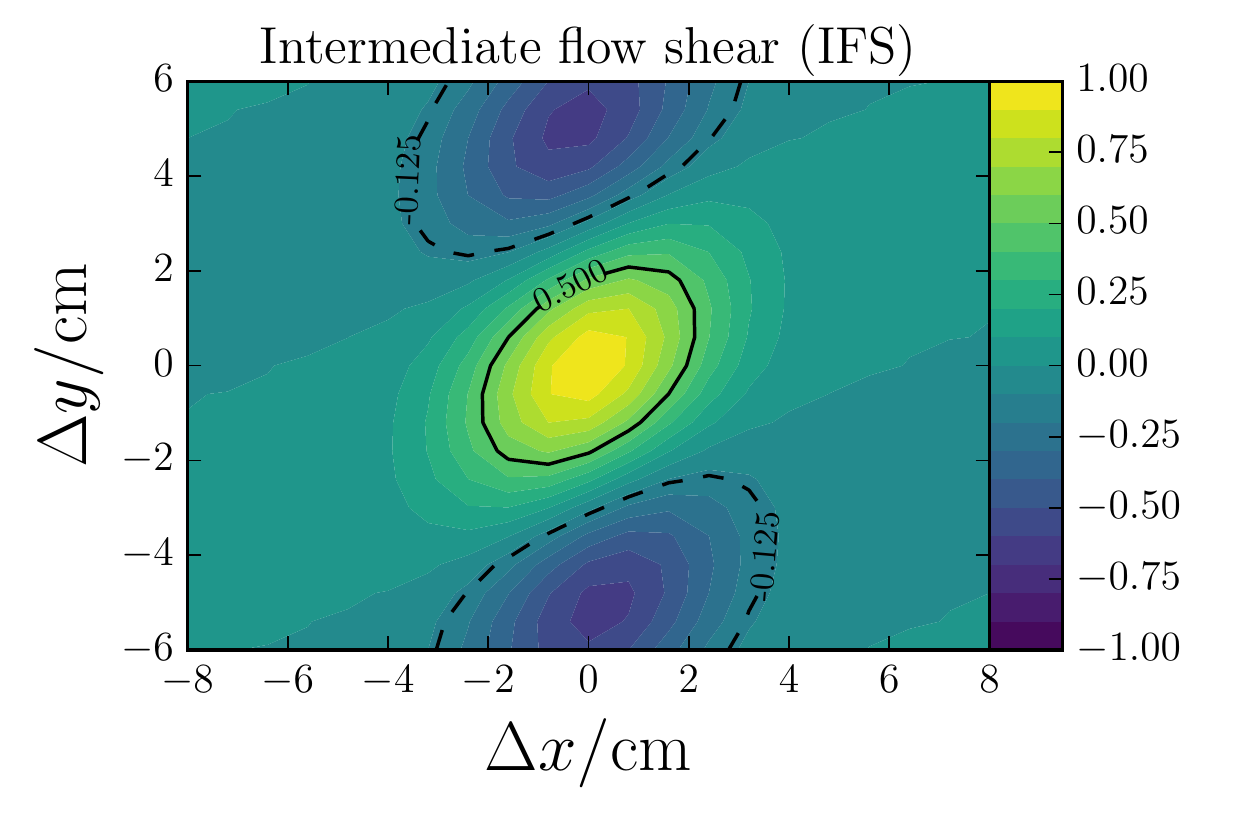}}
\caption{Same as Figure~\ref{fig:compare_LM_shots}, but for the IFS case, whose normalised flow shear (\ref{gammaE_def}) is $\sim40\%$ smaller than in the BLM case---the tilt of the correlation function is also smaller (see Table~\ref{tab:turb_params}). \label{fig:IFS_shot}}
\end{figure}

The measurement uncertainties discussed above and a relatively limited range of values of $\gE$ for which well-resolved BES data is available\footnote{I.e., in the context of this study, the data for which the tilt angle $\Theta$ can be reliably calculated---the limiting factor is, in most cases, the radial resolution, with values of $\ellx$  never very far from the distance between the BES detector channels and from the effective size of the PSFs (see discussion of resolution limits in \cite{Fox2016}).} preclude us from being able to provide here a straightforward $\Theta$ vs.\ $\gE$ parameter study. Generally speaking, larger $\gE$ will give larger $\Theta$, but the spread is large (in Section~\ref{sec:gs2_asymmetry}, we will argue that this is because $\Theta$ is a function not just of $\gE$ but also of how far from the stability threshold the system is). The case of relatively low, but measurable, values of shear is perhaps an interesting benchmark as in this case we expect the tilt to be small. We have found such a case, case IFS, described in Section~\ref{sec:experimental-configuration} and documented in Tables~\ref{tab:equilibria} and~\ref{tab:turb_params}. This has normalised flow shear (\ref{gammaE_def}) that is $\sim40\%$ smaller than that in the BLM case (with its other equilibrium parameters not differing in any particularly remarkable way from the BLM ones).  

The correlation function for the IFS case is shown in Figure~\ref{fig:IFS_shot}. Its tilt is in between that of the BLM and DLM cases and its $\tauc$ is shorter than for the stronger-sheared BLM case (see Table~\ref{tab:turb_params}). Clearly, two points a parameter scan do not make, but this is the best one can do with current data. The salient message is that the transition from stronger flow shear (BLM) to no flow shear (DLM) is gradual in parameter space, with the tilt angle passing through a range of values---and a more comprehensive parameter study ought to be on experimentalists' agenda. The most interesting outcome from such a study would be the dependence of $\tauc$ on $\gE$, telling us how the dynamical properties of turbulence change with shear (cf.~Section~\ref{sec:gs2_away_from_marginal}).   

\subsection{Distribution of fluctuation field}
\label{sec:LM-intensity-dist}

In Section~\ref{sec:intro-skew}, we argued that the breaking of the radial-reflection symmetry by the flow shear can lead to the breaking of the symmetry (evenness) of the distribution of the fluctuating density and, therefore, BES-measured intensity field. Here we show that this is indeed the case. 

We consider the fluctuating part $\delta I_i(t)$ of the intensity field (\ref{Isplit}) and, for each detector channel $i$, normalise it by its own standard deviation 
\begin{equation}
\label{single-channel-std}
\delta I_i^\mathrm{std} = \left[\langle\delta I_i^2(t)\rangle - \langle \delta I_{\mathrm{noise},i}^2\rangle\right]^{1/2},
\end{equation} 
where we have subtracted the mean square amplitude $\langle \delta I_{\mathrm{noise},i}^2\rangle$ of the fluctuating part of the photon noise signal (determined from the signal at each channel during the calibration of the instrument~\cite{Ghim2013}). The angle brackets signify averages over time. The normalisation to $\delta I_i^\mathrm{std}$ is required in order to account for a degree of variation of this quantity across the BES subarray used for our analysis. The total standard deviation (total rms fluctuation amplitude) over all $N=20$ channels in the subarray is
\begin{equation}
\label{std-dev}
\overline{\delta I / I}= \left[\frac{1}{N}\sum_i\left(\frac{\delta I_i^\mathrm{std}}{\langle I_i\rangle}\right)^2\right]^{1/2}.
\end{equation}
This quantity is proportional to the total rms fluctuating density field $\overline{\delta n / n}$, whose value can be reconstructed from it by correcting for PSF effects (see Section 7.1 of \cite{Fox2016}) and is given in Table~\ref{tab:turb_params}. The distribution of the normalised fluctuating intensities $\delta I_i/\delta I_i^\mathrm{std}$ can be affected by several different known experimental effects, including MHD activity, radiation spikes, and PSF effects. A discussion of these effects and how they have been accounted for in our present analysis is given in Appendix~\ref{sec:sources-of-skew}. 

\subsubsection{Symmetry breaking}
\label{sec:sb-skew}

\begin{figure}
\centerline{\includegraphics[width=\textwidth]{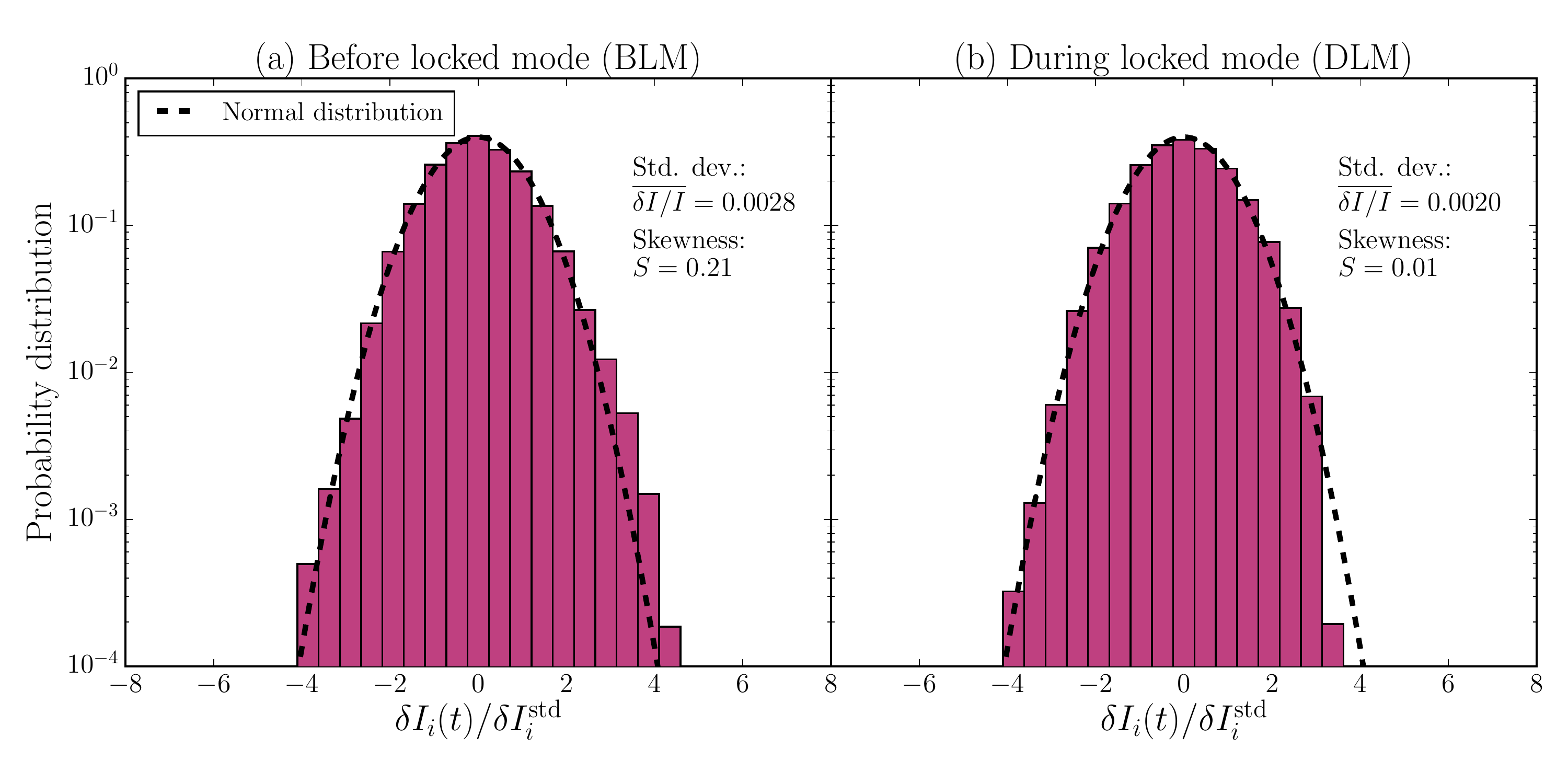}}
\caption{Distribution of the fluctuating intensity field for (a) the case before the locked mode (BLM, with flow shear) and (b) the case during the locked mode (DLM, no flow shear), both described in Section~\ref{sec:experimental-configuration}. For each detector channel, $\delta I_i$ is normalised by its own root-mean-square (standard deviation) value (\ref{single-channel-std}). The black-dashed line in each case gives the unit normal distribution. The distribution function and its skewness for the DLM case (b) were calculated from 2 (rather than 5, as in all other cases) sets of 4 poloidal BES channels, located at $R=1.33$~m and $R=1.35$~m, where the toroidal rotation profile is completely flat, as seen in Figure~\ref{fig:equilibrium_LM}(d). \label{fig:lm_intensity_distribution}}
\end{figure}

In Figure~\ref{fig:lm_intensity_distribution}, we show the probability distribution of the normalised fluctuating intensities $\delta I_i/\delta I_i^\mathrm{std}$ for the BLM and DLM cases. In the case without flow shear (DLM), the distribution is even with respect to positive and negative $\delta I$ (and very nearly normal). In contrast, the case with shear (BLM) exhibits a relatively small, but measurable preponderance of positive $\delta I$, i.e., overdensities are statistically somewhat more common than underdensities. This can be quantified as a positive value of the skewness of the distribution\footnote{The fluctuating part of the photon noise is Gaussian distributed~\cite{Dunai2010,Ghim2012PPCF} and so does not affect the third moment of~$\delta I$.} 
\begin{equation}
\label{skew-def}
S = \frac{1}{N}\sum_i\frac{\langle\delta I_i^3(t)\rangle}{(\delta I_i^\mathrm{std})^3}.
\end{equation}
For the BLM case, $S=0.21$; for the DLM case, $S=0.01$, but this is similar to the skewness in the background emission signal, as well as to the size of the error in the determination of the skewness that occurs due to PSF effects (see Appendix~\ref{sec:app_PSF}), so cannot be considered significant.

\begin{figure}
\centerline{\includegraphics[width=0.5\textwidth]{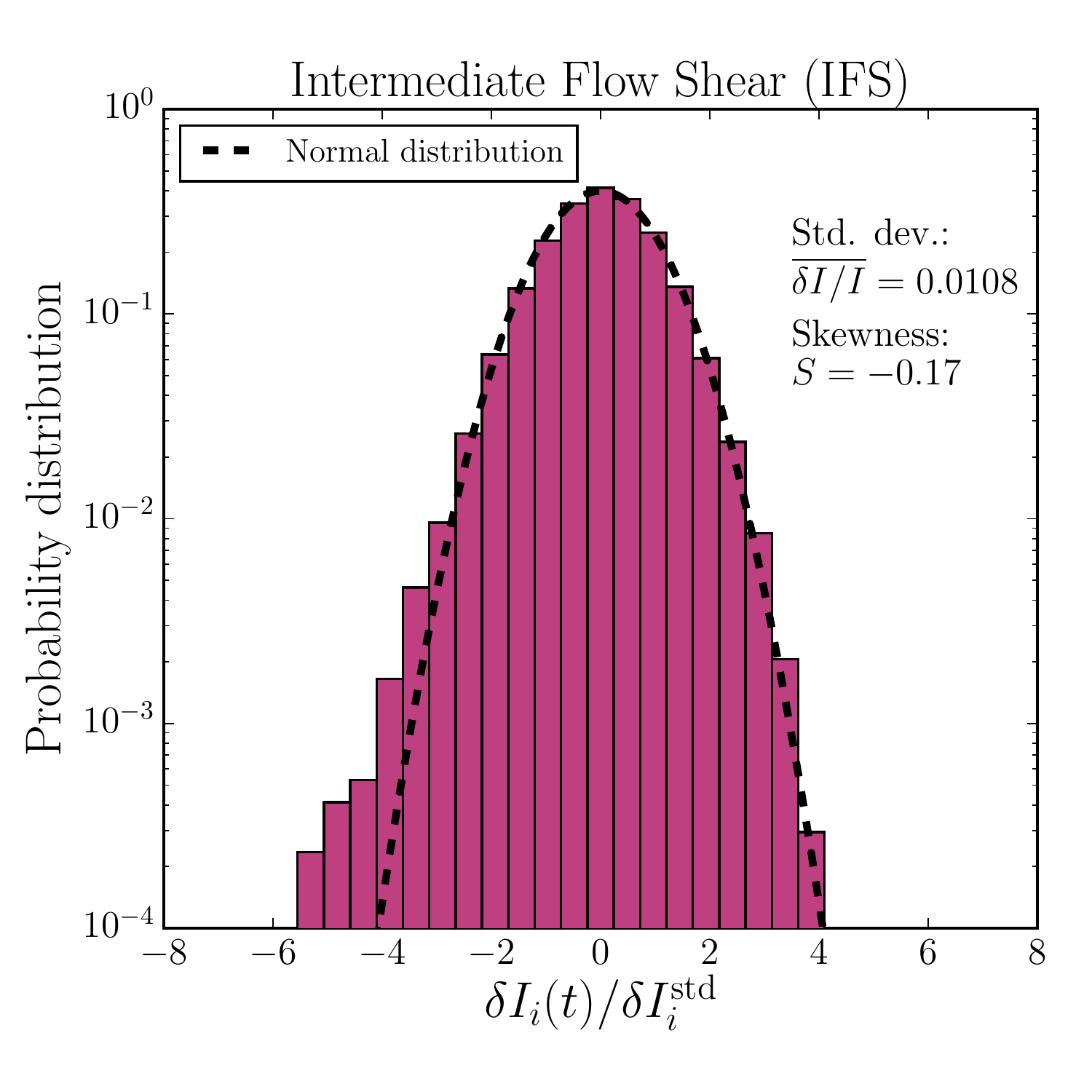}}
\caption{Same as Figure~\ref{fig:lm_intensity_distribution}, but for the IFS case (lower flow shear than for BLM; see Table~\ref{tab:turb_params}). \label{fig:IFS_intensity_distribution} }
\end{figure}

In Figure~\ref{fig:IFS_intensity_distribution}, we show the fluctuating-intensity distribution for the IFS case, 
with flow shear lower than BLM, already discussed in the context of its lower tilt in Section~\ref{sec:IFStilt}. We see that the distribution is again skewed, with its skewness lower than for the BLM case (but still measurable), just as its tilt was (to identify a quantitative dependence between the skewness and flow shear, we will need to resort to simulations, in Section~\ref{sec:gs2-skew}, as we do not have a sufficient range of data for a proper parameter scan). What is more interesting is that the sign of the skew is negative, i.e., here, underdensities are privileged compared to overdensities. This emphasises the fact that, while the presence of a flow shear allows such asymmetric distributions, theory (at least in its current state) does not predict the sign of the asymmetry---we do not know how the sign of the skew depends on local equilibrium parameters. 

Finally, we must stress a very strong caveat. The fact that the symmetry (\ref{symmetry-transforms}) is formally broken by flow shear does not necessarily imply that a skewed distribution should result, it merely allows it (in contrast to the case of zero flow shear, when a skewed distribution would clash with theory \cite{Parra2011,Sugama2011}). Here we have demonstrated, as a proof of principle, that skewed distributions can indeed be found experimentally when flow shear is present. However, skewness is not as robust and inevitable effect of flow shear as tilted correlation functions appear to be and we have found many experimental cases in which flow shear is present according to CXRS measurements but the skewness of the fluctuating-intensity distribution is so small that, in view of measurement uncertainties, it cannot be declared present beyond reasonable doubt. In Section~\ref{sec:gs2_asymmetry}, we will see, by analysing numerical simulations, that how skewed the distribution is depends quite strongly on how far from the nonlinear stability threshold the system finds itself, with only near-marginal cases showing significant skewness. Thus, while a tilt (possibly small) in the correlation function might be viewed as a signature of the presence of flow shear (as argued in Section~\ref{sec:corr_times}), a noticeable skewness of the fluctuating-density distribution should perhaps be viewed as a signature of a sheared turbulent state being close to threshold. The physical reason for this is probably that, close to the threshold, the turbulent state in MAST is dominated (as shown in \cite{vanWyk2016} and discussed in Section~\ref{sec:physics-skew}) by a finite number of long-lived, finite-amplitude coherent structures, whereas farther from the threshold, a more (evenly) distributed sea of fluctuations emerges.   

\subsubsection{Skewness vs.\ tilt: two-component turbulence?} 
\label{sec:core_tail}

\begin{figure}
\centerline{\includegraphics[width=\textwidth]{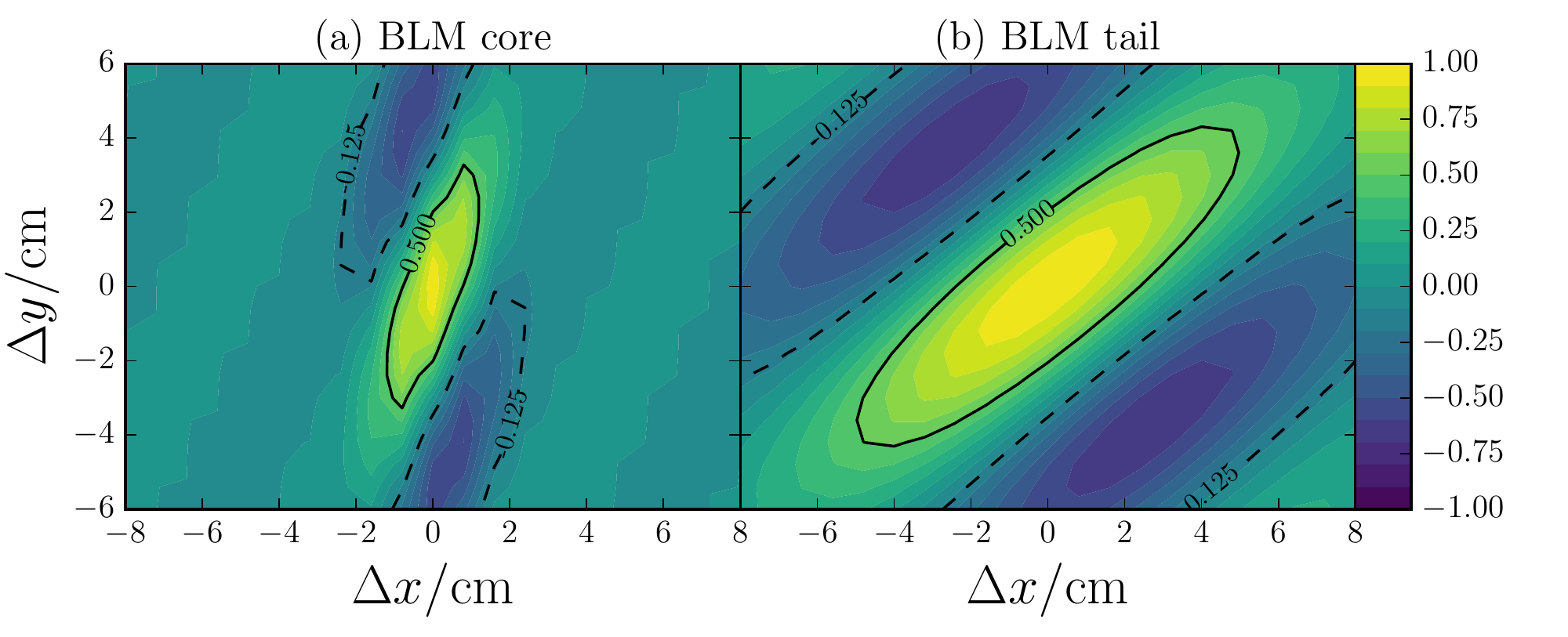}}
\caption{Same correlation function as in Figure~\ref{fig:compare_LM_shots}(a), for the BLM case, but this time conditioned on (a) lower intensities (core of the distribution, $|\delta I_i/\delta I_i^\mathrm{std}| < 2.75$) and (b) higher intensities (tail of the distribution, $\max \delta I_i/\delta I_i^\mathrm{std} > 3$), in the way described in Section~\ref{sec:core_tail}. The parameters of the fitting function (\ref{plasma-corr-fun}) in each case are given in Table~\ref{tab:core_tail}. The tail is manifestly less tilted than the core and also has a larger radial correlation length. These correlation functions should be compared to the ones for numerically simulated turbulence in Figure~\ref{fig:gs2-conditional}. \label{fig:BLM_core_tail} }
\end{figure}

The above discussion and the greater robustness of the tilted correlation functions than of the skewed distributions as signatures of flow shear suggest that these features describe the response to the flow shear of different types of turbulent fluctuations. Indeed, intuitively, as perturbations are sheared by the flow, their radial wave number increases according to (\ref{kx_shear}) and so then does $k_\perp\rho_i = (\kx^2 + \ky^2)^{1/2}\rho_i$. This will push them into a damped region of wave-number space (see, e.g., \cite{Schekochihin2012}), quickening their demise. It is these perturbations, which are on their way out, that would statistically contribute to the tilt in the correlation function. In contrast, one might imagine that rare but particularly strong perturbations whose presence skews the distribution of the fluctuating field, might be less vulnerable to (or simply not yet affected by) such destruction by shear and so would turn out to be less tilted. This view is perhaps reinforced by the observation, which we will be in a position to make when we consider the numerically simulated turbulence in Section~\ref{sec:gs2-skew}, that the rms fluctuation level $\overline{\delta n/n}$ is significantly lower in the turbulence with flow shear than without it and so the tail of the distribution represents not some additional population appearing alongside the ``core turbulence'' but rather the ``survivors''---sufficiently large structures that can create sufficiently large local velocity gradients to overcome the suppressing effect of the shear; these large structures would diffuse and disappear rapidly in the strong unsheared turbulence.

These speculations can be experimentally tested, if in a rather tentative manner, by constructing a correlation function conditional on the fluctuation amplitude. Indeed, the turbulent density field is continuously (and quickly) advected by the flow past our, relatively small, BES subarray, which can thus be thought of as sampling different types of turbulent structures at different times, rather than capturing a fully statistically representative spatial domain containing all types of structures at all times. Thus, we collect all snapshots of the turbulent field (such as those discussed in Section~\ref{sec:bes_2d_frames}) in which the fluctuating intensities at all detector channels are below a certain threshold, specifically, $|\delta I_i/\delta I_i^\mathrm{std}| < 2.75$, and calculate the spatial correlation function of this collection, representing the ``core'' of the distribution, in the way described in Section~\ref{sec:BES-spatial-correlation-function}. We repeat the same procedure for a complementary set of snapshots, in which at least one detector channel sees $\delta I_i/\delta I_i^\mathrm{std} > 3$, to extract the correlation parameters of the ``tail'' of the distribution. This analysis is necessarily crude, because the reduced number of snapshots (especially for the tail) reduces the quality of the statistics and the size of the BES window may have an effect that we cannot study systematically. Proceeding with it nevertheless, we find that the tilt angle for the core of the distribution in the BLM case is noticeably larger than in the tail (and also that the radial correlation length $\ellx$ in the tail is much larger than in the core; see Table~\ref{tab:core_tail}). For the DLM case, there is no significant difference between the tail and the core, the tilt angle is essentially zero for both. This result appears to support the line of reasoning proposed above, although obviously a much more extensive study of a large number of well-resolved cases with good statistics is called for.

The need for such a study to make our argument fully convincing (or otherwise) is accentuated by the disconcerting failure of the IFS case to agree with the core vs.\ tail distinction seen in the BLM case: in fact, we find that the tilt angle in the tail is larger than in the core (see Table~\ref{tab:core_tail}). We do not know why the IFS case has this behaviour: whether, for example, it might be related to the lower (dynamically unimportant?) flow shear, or to the preponderance of under-, rather than overdensities in the distribution of the fluctuating field, or to the presence of zonal flows.\footnote{As discussed in Section~\ref{sec:ZF}, turbulence in MAST is expected (based on numerical simulations) to develop both zonal flows and zonal density perturbations. Locally, zonal shears will cause correlation functions to become tilted in the same way as mean flow shear does. Depending on the life time of the zonal fields and on the high-pass frequency filter (i.e., effectively, on the definition of time average in (\ref{Isplit})) applied to the intensity signal (see Appendix~\ref{sec:correct_skew}), we have found numerically that the presence of a residual zonal density perturbation in the signal can bias conditional correlation functions calculated for stronger or weaker intensities towards regions of positive or negative zonal densities, which are also regions of stronger zonal shear. One can imagine that, when the mean flow shear is weak, this could matter and lead to greater tilt angles in the tail of the distribution, as found here for the IFS case.} In Section~\ref{sec:skew-tilt}, we will see that the trend in numerical simulations appears to be broadly consistent with the BLM case. 

\begin{table} 
	\centering 
	\begin{tabular}{r|ll|ll} 
		Case & BLM core& BLM tail& IFS core& IFS tail \\
                Condition & $|\delta I_i/\delta I_i^\mathrm{std}|<2.75$ & $\delta I_i/\delta I_i^\mathrm{std}>3$ & $|\delta I_i/\delta I_i^\mathrm{std}|<2.75$ & $\delta I_i/\delta I_i^\mathrm{std}<-3$ \\ 
		No.\ of snapshots & 1434 & 123 & 3065 & 206 \\ 
                \hline\rule{0pt}{1.2em}
		$\ellx/\rho_i$ & $1.9 \pm 0.2$ & $ 8.4 \pm 0.3$ & $2.9 \pm 0.1$ & $4.8 \pm 0.3$ \\
		$\elly/\rho_i$ & $9.2 \pm 0.1$ & $ 9.2 \pm 0.3$ & $10.4 \pm 0.1$ & $11.0 \pm 0.2$ \\
		$\kx\rho_i$    & $-1.00 \pm 0.14$ & $-0.36 \pm 0.01$ & $-0.18 \pm 0.02$ & $-0.35 \pm 0.02$\\
		$\ky\rho_i$    & $ 0.44 \pm 0.01$ & $ 0.43 \pm 0.01$ & $0.46 \pm 0.1$ & $0.43 \pm 0.01$\\ 
		Tilt: $\Theta$/deg & $67 \pm 3$ & $40\pm1$ & $22\pm3$ & $39\pm2$ 
	\end{tabular}
	\caption{Correlation parameters of the fitting function (\ref{plasma-corr-fun}) for the BLM and IFS core and tail conditional correlation functions (for the BLM case, they are shown in Figure~\ref{fig:BLM_core_tail}). The errors given here are for the fitting procedure only. The parameters for the overall correlation functions can be found in Table~\ref{tab:turb_params}. Note that the number of snapshots does not represent the number of statistically independent instances: neighbouring snapshots are separated by $0.5~\mu\mathrm{s}$, so it takes approximately 20 snapshots for a given structure to pass through the BES subarray (due to the apparent poloidal motion caused by toroidal rotation \cite{Ghim2012PPCF}). \label{tab:core_tail}} 
\end{table}

Cognizant of all the uncertainties and of the clearly insufficient amount of experimental evidence on offer, we conclude this section with a cautious conjecture that turbulence under the action of flow shear consists of a tilted, sheared, statistically more ubiquitous, but smaller-amplitude component and less tilted, rarer, but stronger perturbations, which skew the distribution function of the fluctuating field. We shall test this idea further in the next section. 

\section{Symmetry breaking in numerically simulated gyrokinetic turbulence}
\label{sec:gksims}

Having established experimentally, at least as a proof of principle, that the presence of the flow shear breaks both the spatial symmetry of the turbulence and the symmetry of the distribution of the fluctuating density perturbations, we now turn to gyrokinetic simulations, both to check that modelling can reproduce what theory predicts and the experiment has revealed and to seek clearer guidance on the parameter dependence of these effects. These numerical results give us a degree of confidence in the validity of our interpretation of the experimental data, which, especially in what concerns the skewed distributions (Section~\ref{sec:LM-intensity-dist}), was perhaps not on its own sufficient to build a rock-solid case.    

In this section, we use a set of local, flux-tube, electrostatic, two-species, ion-scale numerical simulations using the gyrokinetic code GS2\footnote{http://ter.ps/gs2} that were originally performed for a different, independent study of the transition to turbulence in MAST \cite{vanWyk2016}. The experimental case whose equilibrium parameters served as input for these simulations has previously been used for validation and verification exercises involving both NEMORB \cite{FieldPPCF2014} and GS2 \cite{vanWyk2016b}. This experimental case is labelled EGK and its parameters are given in Table~\ref{tab:equilibria}. While the BES data from this case has been studied extensively using 1D reconstruction of various correlation parameters \cite{FieldPPCF2014}, it unfortunately turns out not to be suitable for a full-scale 2D analysis as developed in \cite{Fox2016}, because calculation of $\kx$ and, therefore, the tilt, suffers from insufficient resolution. However, it is not wildly different from our BLM case\footnote{It does have a substantially larger flow shear, which was the reason for all the numerical interest in it in the first place, but, in view of the arguments advanced below, the important parameter is not the absolute value of $\gEhat$ by itself but the distance to the stability threshold (see Section~\ref{sec:gs2_asymmetry}).} and, in any event, our aim here is to determine qualitative trends rather than to claim precise agreement between simulations and experiment (although a more limited 1D analysis of the BES data for the EGK case does show respectable agreement with GS2 results \cite{vanWyk2016b}). 

The obvious advantage accorded by (local flux-tube) simulations is that one can change individual equilibrium parameters at will, while keeping other parameters constant, and thus look for trends. We will take this route by analysing a sequence of simulations in which the flow shear $\gEhat$ is varied compared to the nominal EGK case---and confirm that the gyrokinetic simulations are able to recover, qualitatively, both the tilted correlation functions (Section~\ref{sec:gs2_symmetry_breaking}) and the skewed distributions (Section~\ref{sec:gs2-skew}) observed in the experiment. 

\begin{table}
\centering
\begin{tabular}{r|rrrrrr} 
Run name &GKa5&GKa4&GKa3&GKa2&GKa1&Marginal\\ 
$\gEhat$& $0.00$& $0.08$& $0.10$& $0.12$& $0.14$& $0.16$\\ 
\hline\rule{0pt}{1.2em}
$\Qgb$& $35.1$& $24.6$& $18.0$& $13.0$& $13.0$& $1.4$\\ 
$\delta n/n$& $0.073$& $0.056$& $0.044$& $0.040$& $0.034$& $0.011$\\ 
Skewness& $-0.055$& $0.0057$& $0.097$& $0.053$& $0.22$& $2.3$\\ 
\hline\rule{0pt}{1.2em}
$\ellx/\rho_i$& $4.6$& $3.8$& $3.9$& $3.5$& $3.8$& $3.5$\\ 
$\elly/\rho_i$& $10$& $8.9$& $8.7$& $8.8$& $8.8$& $10$\\ 
$\kx\rho_i$& $-0.0057$& $-0.16$& $-0.23$& $-0.22$& $-0.37$& $-0.51$\\ 
$\ky \rho_i$& $0.26$& $0.29$& $0.29$& $0.28$& $0.27$& $0.24$\\ 
Tilt $\Theta/\mathrm{deg}$& $1.5$& $35$& $44$& $44$& $59$& $69$\\ 
\hline\rule{0pt}{1.2em}
$\tauc$& $-$& $8.6$& $9.6$& $7.9$& $12$& $16$\\ 
\hline\rule{0pt}{1.2em} 
CORE &&&&&\\
$\ell_x/\rho_i$& $3.7$& $3.3$& $3.0$& $3.0$& $3.3$& $2.7$\\ 
$\ell_y/\rho_i$& $8.9$& $7.8$& $7.7$& $7.8$& $7.7$& $9.3$\\ 
$k_x\rho_i$& $0.012$ & $-0.19$& $-0.25$& $-0.28$& $-0.41$& $-0.56$\\ 
$k_y \rho_i$& $0.27$& $0.25$& $0.29$& $0.29$& $0.27$& $0.18$\\ 
Tilt $\Theta/\mathrm{deg}$& $-3.2$& $39$& $46$& $50$& $62$& $75$\\ 
\hline\rule{0pt}{1.2em} 
TAIL &&&&&\\
$\ell_x/\rho_i$& $5.1$& $4.4$& $4.0$& $4.1$& $4.5$& $3.6$\\ 
$\ell_y/\rho_i$& $12$& $9.8$& $10$& $10$& $9.5$& $10$\\ 
$k_x\rho_i$& $-0.020$& $-0.14$& $-0.14$& $-0.17$& $-0.35$& $-0.50$\\ 
$k_y \rho_i$& $0.28$& $0.30$& $0.30$& $0.29$& $0.28$& $0.26$\\ 
Tilt $\Theta/\mathrm{deg}$& $5.1$& $29$& $30$& $35$& $56$& $67$\\ 
\end{tabular} 
\caption{Various statistical characteristics of turbulence in a series of simulations corresponding to the equilibrium parameters of the EGK case (Table~\ref{tab:equilibria}, except $a/L_{T_i}=4.8$ in these runs) and a sequence of values of flow shear $\gEhat$. The Marginal case is the EGK case itself. The rms fluctuation amplitude $\overline{\delta n/n}$ is calculated using (\ref{std-dev}), but replacing $\delta I_i(t)$ with $\delta n_i(t)$; the summation is over all grid points~$i$. Similarly, the skewness is calculated for the fluctuating density field using (\ref{skew-def}). The spatial correlation functions for the GKa5, GKa3, GKa1, and Marginal cases are plotted in Figure~\ref{fig:gs2_compare_correlation_functions}. The fluctuating-density distributions for the GKa5, GKa1, and Marginal runs (the latter two skewed) are shown in Figure~\ref{fig:gs2_distributions}. The conditional correlation functions (CORE and TAIL) are discussed in Section~\ref{sec:skew-tilt} and plotted in Figures~\ref{fig:gs2-conditional} (run GKa1) and \ref{fig:gs2-conditional-marg} (Marginal run). \label{tab:flow_shear_scan}}
\end{table}

\begin{figure}
\centerline{\includegraphics[width=\textwidth]{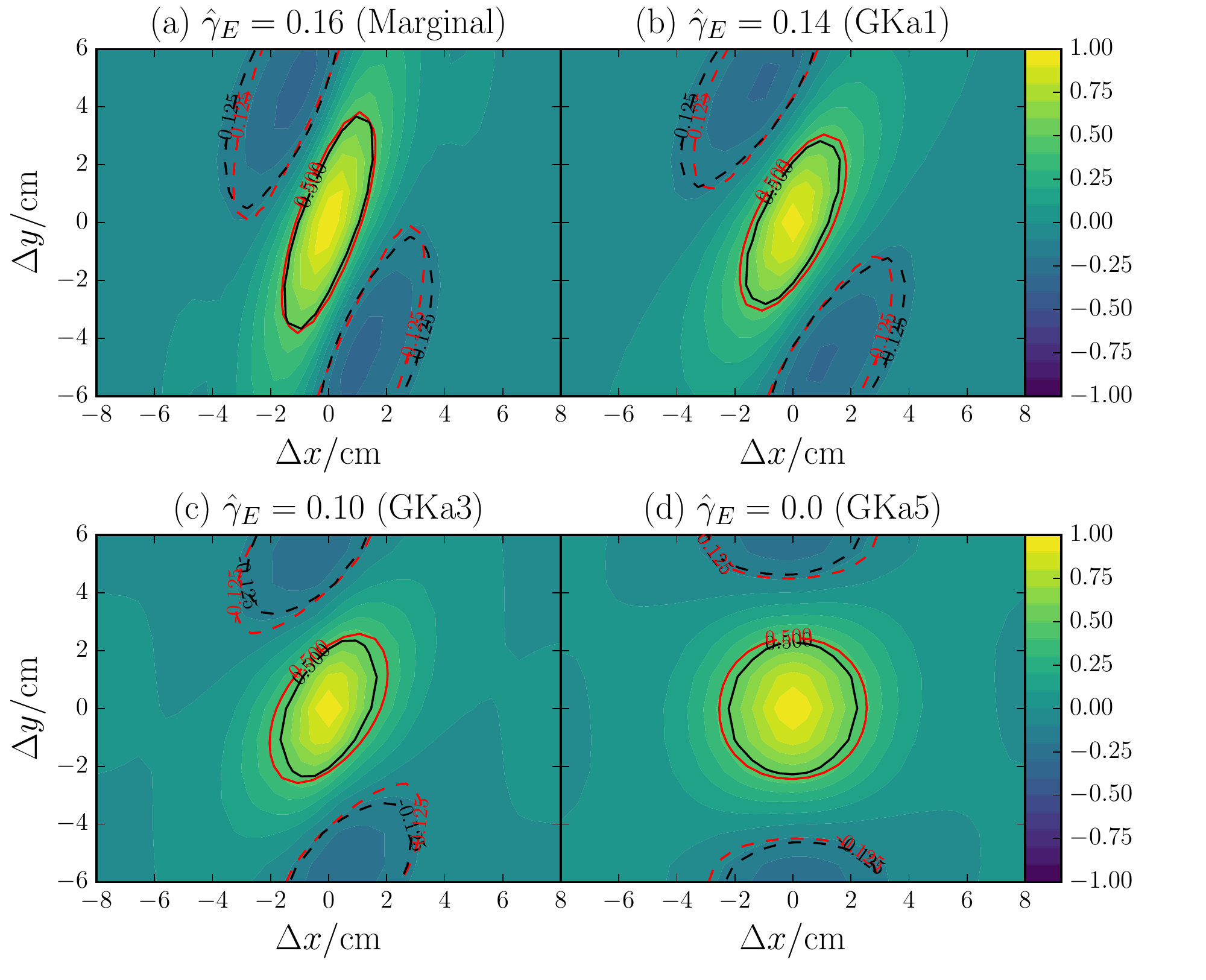}}
\caption{Directly calculated spatial correlation functions of the fluctuating density field for 4 of the runs documented in Table~\ref{tab:flow_shear_scan}, with, from (a) to (d), value of the flow shear decreasing from the experimental (EGK) value to $\gEhat=0$. The spatial domain of the simulation is regularly gridded, therefore multiple values of the correlation function (\ref{correlation_def}) (with $\delta n_i$ instead of $\delta I_i$, $i$ being the grid point) occur for each $(\dr, \dZ)$ pair; these values are then averaged to produce the spatial correlation function that is plotted (it is also averaged over time, typically several hundred $\mu\mathrm{s}$, i.e., tens of correlation times). These correlation functions are to be compared with experimental correlation functions with and without flow shear in Figure~\ref{fig:compare_LM_shots}. The correlation parameters for the fitting function (\ref{plasma-corr-fun}), approximating the true correlation functions plotted here, are given in Table~\ref{tab:flow_shear_scan}. The red contours in these plots correspond to the fitting function (\ref{plasma-corr-fun}) with these parameters, showing the quality of the fit (and thus supporting its use for experimental data; see \cite{Fox2016} for further discussion). \label{fig:gs2_compare_correlation_functions}}
\end{figure}

\begin{figure}
\centering
\begin{tabular}{cc}
\includegraphics[width=0.5\textwidth]{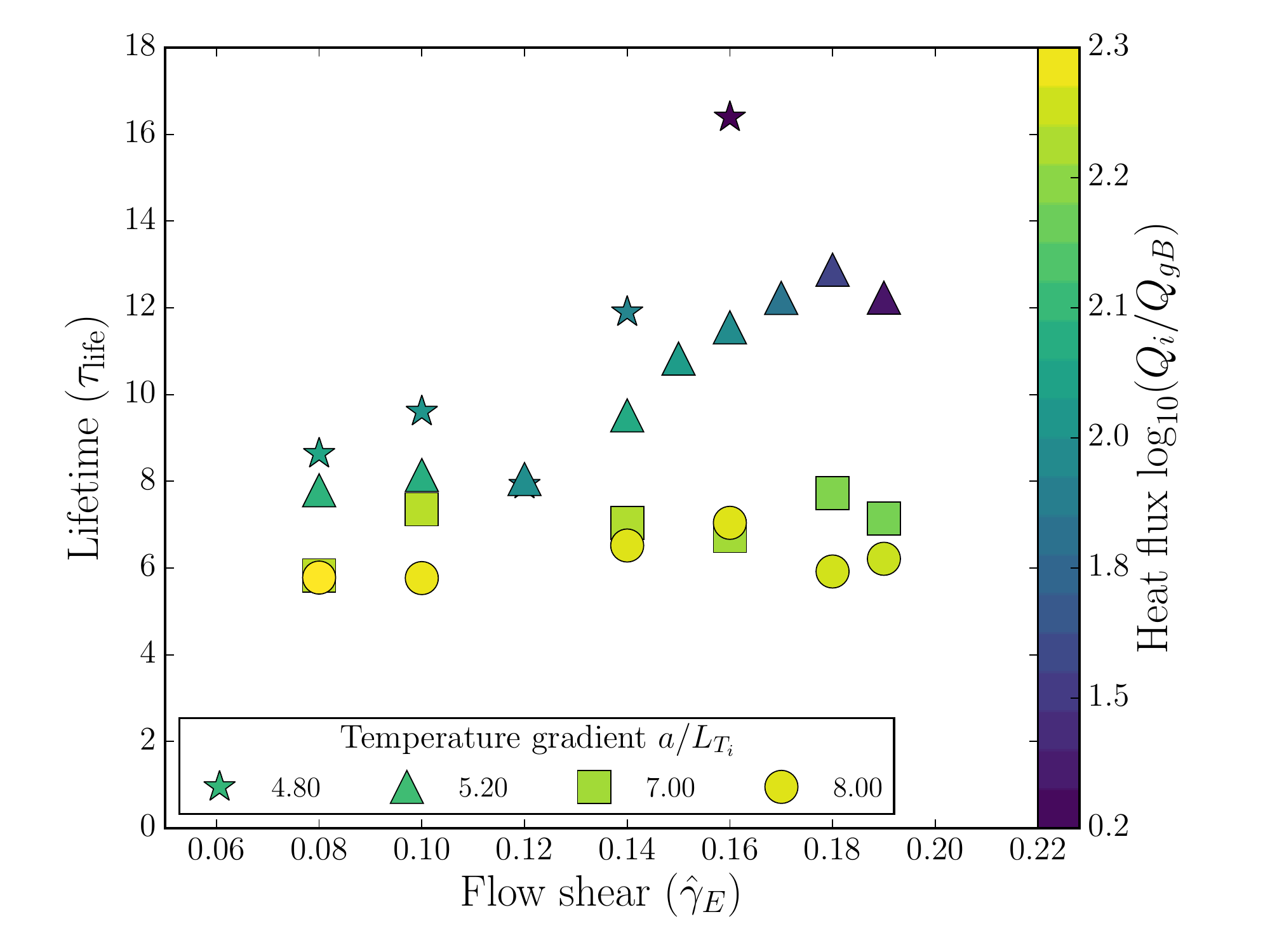} & 
\includegraphics[width=0.5\textwidth]{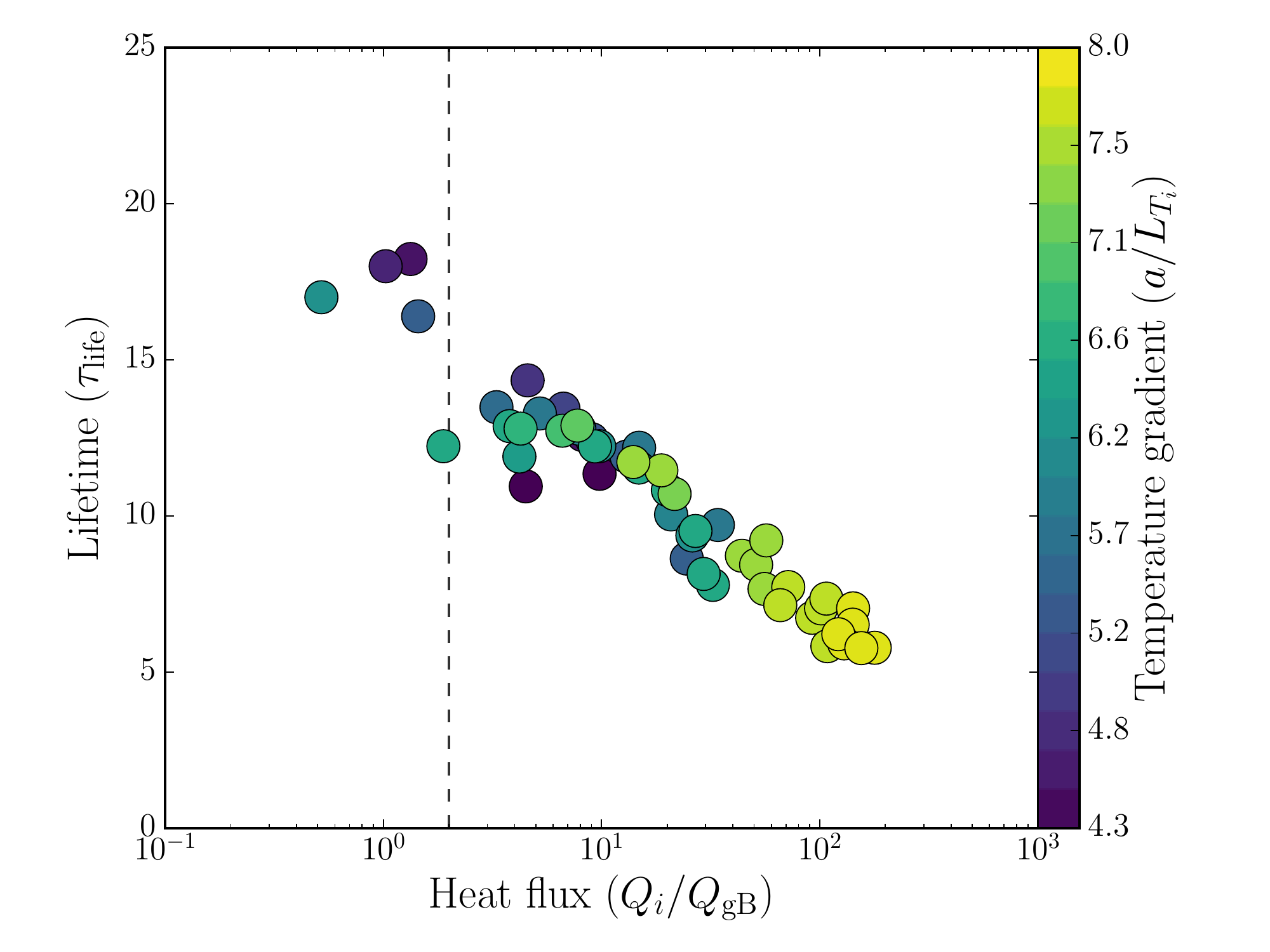}\\
(a) $\tauc$ vs.\ $\gEhat$ & (b) $\tauc$ vs.\ $\Qgb$
\end{tabular}
\caption{(a) Life time of perturbations, defined by (\ref{tauc-def}), $\tauc = |\tan\Theta/\gE|$, and normalised by $a/\vthi$, vs.\ flow shear $\gEhat$ for a number of values of the ion-temperature gradient $a/L_{T_i}$ (distinguished by different shapes of the data points); the data points are coloured according to the value of the gyro-Bohm-normalised heat flux $\Qgb$ (see discussion in Section~\ref{sec:gs2_asymmetry}). (b) Same, but here $\tauc$ is plotted vs.\ $\Qgb$ with data points coloured according to the value of $a/L_{T_i}$ (this plot includes data from the full set of simulations carried out in \cite{vanWyk2016} and so covers a larger number of values of ITG than (a)); the vertical dashed line indicates the experimental value of $\Qgb$ for the EGK case (Table~\ref{tab:equilibria}). \label{fig:taulife_vs_gE}}
\end{figure}

However, it must be clear that this does not, in fact, constitute a parameter scan that one might expect to achieve experimentally in a real plasma. Indeed, the suppression of the turbulence caused by increasing the flow shear will cause the plasma equilibrium to respond: as the turbulent transport is reduced, assuming constant ion heat flux, the ion-temperature gradient (ITG) must increase~\cite{Highcock2012,Ghim2014}. Generally speaking, experimentally one expects to find turbulent plasma states near the stability threshold (which, in the instances of ion-scale MAST turbulence that we are considering is a nonlinear stability threshold, the turbulence being subcritical \cite{vanWyk2016}). As a crude simplification, we can think of this threshold (or ``zero-turbulence manifold'' \cite{Highcock2012}) as a boundary in a two-dimensional parameter space consisting of $\gEhat$ and ITG, although obviously other equilibrium parameters are in fact also involved \cite{Newton2010,Highcock2012,Ghim2014,Highcock2016}. From the point of view of interpreting experimental results, the key questions are (i) near which part of the zero-turbulence manifold the experimental case under consideration is located and (ii) how close to the threshold it is. The second question is very crucial indeed, because we know from simulations \cite{vanWyk2016} (and will further confirm in what follows) that the properties of our turbulence depend on the distance from threshold very sensitively. To investigate these matters, in Section~\ref{sec:gs2_asymmetry}, we will consider a set of numerical simulations in a range of values of both flow shear and ITG and argue that both the tilt and the skewness are strong functions of the distance from threshold, i.e., that the symmetry-breaking effects associated with the flow shear fade away and the symmetry is restored if one probes deep into the strongly turbulent regime far from the stability boundary. 

\subsection{Tilt vs.\  flow shear}
\label{sec:gs2_symmetry_breaking}

We start by considering a parameter scan in flow shear, with 6 different values of $\gEhat$ (see Table~\ref{tab:flow_shear_scan}) and all other equilibrium parameters the same as those of the EGK experimental case, listed in Table~\ref{tab:equilibria} (except, in these runs, $a/L_{T_i}=4.8$, which is within the experimental uncertainty of the measurement \cite{vanWyk2016}). The largest value of the flow shear in this scan, $\gEhat=0.16$, corresponds to the EGK case itself---and at even larger values, there is no turbulence (all perturbations decay), so we will refer to the simulated EGK case as Marginal.  

For the purposes of this analysis, we adopt the definition of the fluctuating density field $\delta n_i(t)$ analogous to the definition (\ref{Isplit}) of the fluctuating intensity $\delta I_i(t)$ in the experimental measurements reported in Section~\ref{sec:experiment1}: at each grid point $i$, it is the density perturbation at this point (extracted directly from the simulations---it is the velocity integral of the perturbed particle distribution function, which is being solved for by GS2), with a running time average, calculated over a moving $50~\mu\mathrm{s}$ interval, subtracted, so the time average of the fluctuating field is zero by definition. Note that this corresponds to using a $20$~kHz high-pass frequency filter (which experimentally was needed anyway; see Appendix~\ref{sec:correct_skew}) and that this procedure will remove any long-lived zonal density component (see Section~\ref{sec:ZF}). 

The spatial correlation functions of the fluctuating density fields calculated in this fashion are shown in Figure~\ref{fig:gs2_compare_correlation_functions} and the correlation parameters (lengths and wave numbers) defined by the fitting function (\ref{plasma-corr-fun})\footnote{The simulations obviously have a much higher spatial resolution than the BES diagnostic, with many more grid points than BES has channels. Therefore, unlike in Section~\ref{sec:BES-spatial-correlation-function}, we do not need to constrain the product $\kZ\ell_Z$ with the aid of the time-correlation function.} are documented in Table~\ref{tab:flow_shear_scan}. It is manifest that, as the flow shear increases, the radial-reflection symmetry is broken and the correlation function develops a tilt, which becomes quite large as the Marginal case is approached. Qualitatively, the difference between the zero-flow-shear, untilted correlation function for run GKa5 and the sheared, tilted ones for runs GKa3 or GKa1 is very similar to the difference between the DLM and BLM cases shown in Figure~\ref{fig:compare_LM_shots}. 

\subsubsection{Tilt and life time}
\label{sec:tilt-taulife} 

Quantitatively, we find that the tilt $\tan\Theta = \kx/\ky$ increases approximately linearly with $\gEhat$, except near the stability threshold. Recalling the discussion at the end of Section~\ref{sec:intro-tilt} and in Section~\ref{sec:corr_times}, we conclude that, except near the stability threshold, the life time (\ref{tauc-def}) of the turbulent perturbations is independent of the flow shear---presumably because dynamically the flow shear is not very important away from the threshold. This point is illustrated by Figure~\ref{fig:taulife_vs_gE}(a). The data for other values of the ion temperature gradient also shown in this figure supports this conclusion and will be further discussed in Section~\ref{sec:gs2_away_from_marginal}. 

\subsubsection{Tilts at zero flow shear as signature of zonal flows}
\label{sec:ZF} 

\begin{figure}
\centering
\begin{tabular}{cc}
\includegraphics[width=0.45\textwidth]{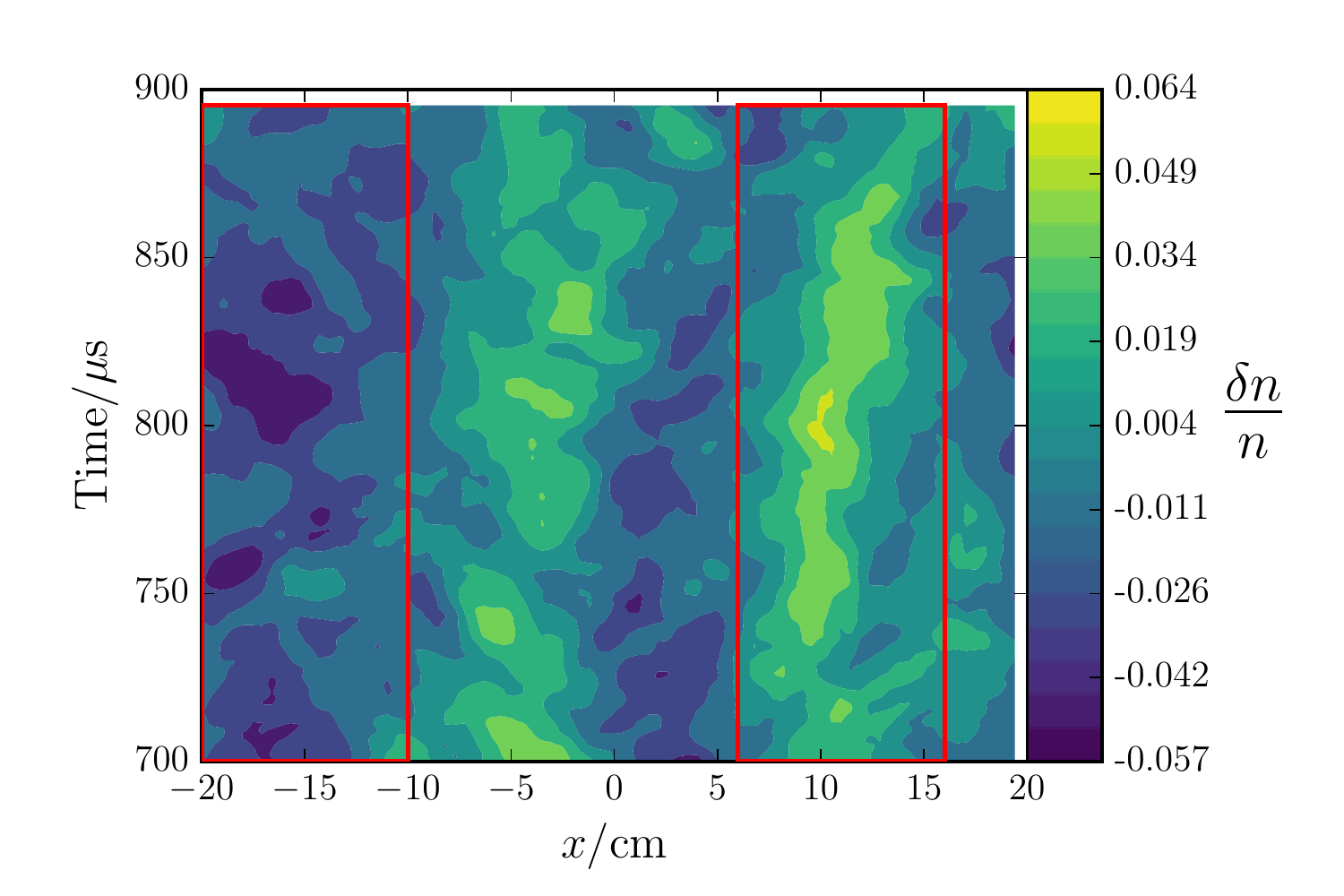} &
\includegraphics[width=0.45\textwidth]{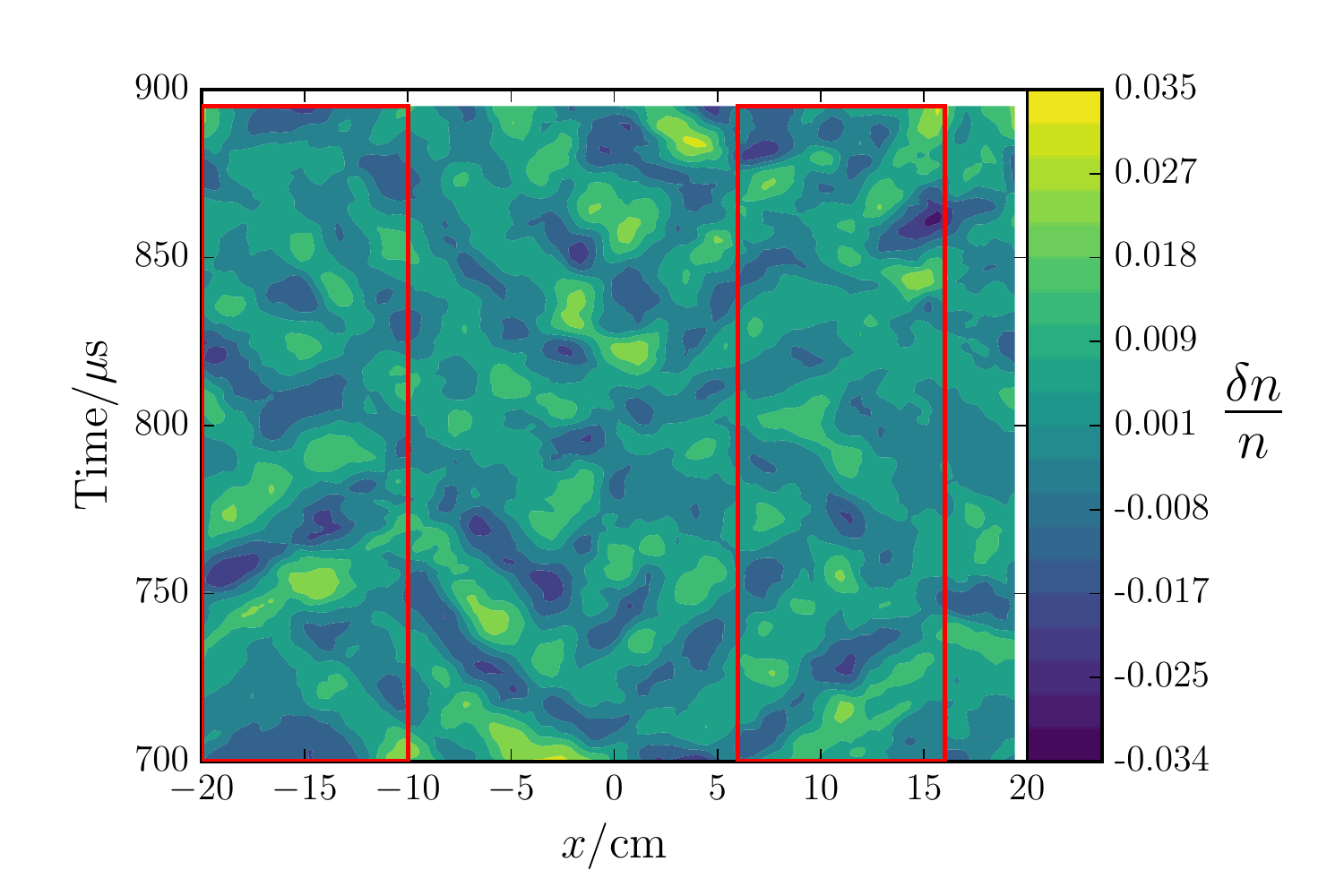} \\
(a) without high-pass filter & (b) with high-pass filter
\end{tabular}
\caption{Perturbed density field from run GKa5 ($\gEhat=0$) integrated over $y$ shown vs.\ $x$ and time (a) without subtracting moving time average and (b) with moving time average subtracted (as for all fluctuating density fields used in Section~\ref{sec:gksims}). The presence of a zonal density component is manifest. The red boxes show regions for which conditional correlation functions shown in Figure~\ref{fig:corr-fn-ZF} and discussed in Section~\ref{sec:ZF} were calculated. \label{fig:ZF}}
\end{figure}

\begin{figure}
\centerline{\includegraphics[width=\textwidth]{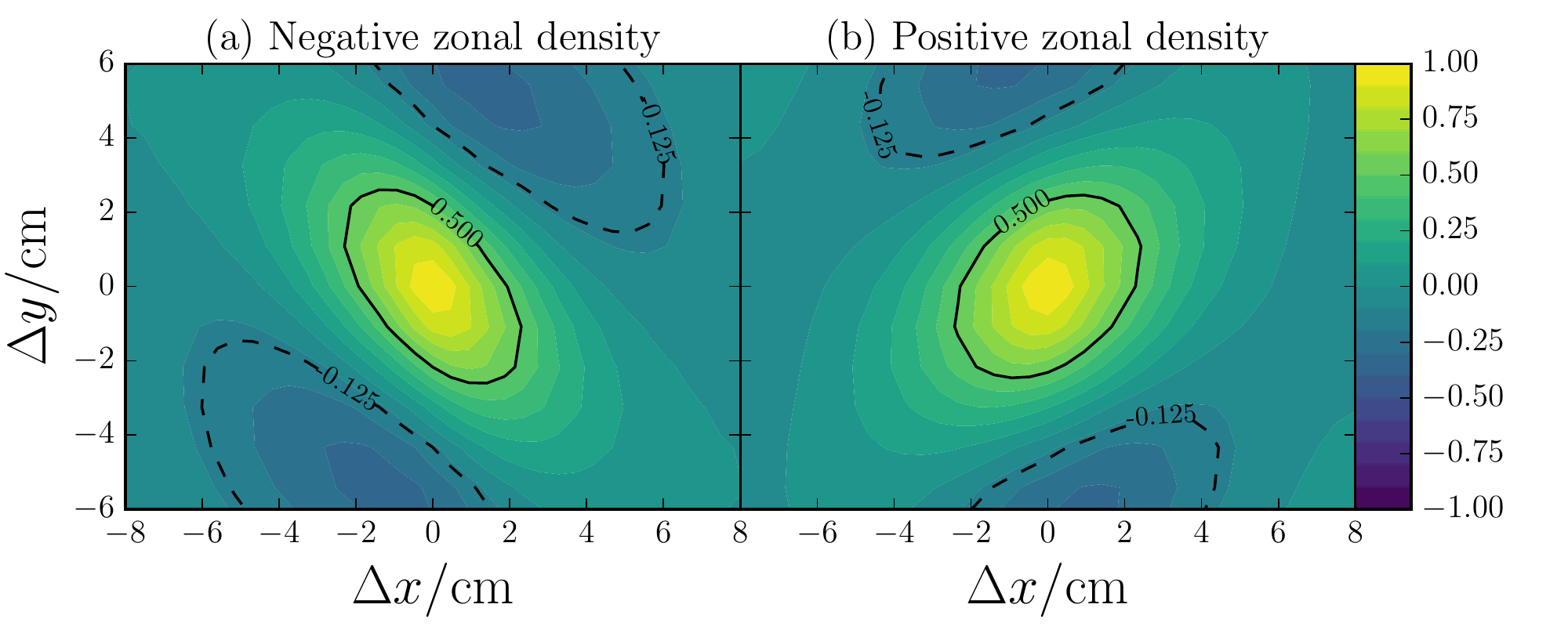}}
\caption{Same correlation function as Figure~\ref{fig:gs2_compare_correlation_functions}(d), for run GKa5 ($\gEhat=0$), but calculated for two spatial subdomains: (a) with positive zonal density and (b) with negative zonal density, as discussed in Section~\ref{sec:ZF}. \label{fig:corr-fn-ZF}}
\end{figure}

As we saw in Figure~\ref{fig:gs2_compare_correlation_functions}(d), when the mean flow shear is zero, the correlation function of the fluctuating density field is untilted, in agreement with the symmetry~(\ref{symmetry-transforms}). We stress, however, that this is only true if the correlation function is calculated over a sufficiently large spatial domain. Because drift-wave turbulence is prone to developing zonal flows \cite{Rogers2000,Diamond2005}, locally a flow shear can be present---and indeed is, in the simulations that we are analysing. As these simulations have kinetic electrons, they can, and do, develop a zonal density perturbation on ion scales, so the presence of a zonal field is detectable directly from the measured density field, in the form of a $y$-independent, long-lived component. Because zonal fields have a long life time, compared to the life time of the turbulence, they can only be seen if we remove the high-pass frequency filter effectively imposed on our fluctuating density field by subtracting the running time average (as stated at the beginning of Section~\ref{sec:gs2_symmetry_breaking}). This is illustrated in Figure~\ref{fig:ZF}, where we show the $y$-integrated perturbed density field for the zero-flow-shear run GKa5 with and without this filter.  

Experimentally, the high-pass frequency filter may be difficult to abandon (see Appendix~\ref{sec:correct_skew}), however, the zonal shear can be detected indirectly via correlation functions of the fluctuating density field:\footnote{Since zonal flows advect the fluctuating density field, it is also possible to detect them by tracking the movement of density contours in time \cite{FieldIAEA2014}.} if those are calculated over a spatial domain that is smaller (radially) than the wave length of the zonal flow, they will be tilted by the same mechanism as in the case of a mean flow shear. This is illustrated in Figure~\ref{fig:corr-fn-ZF}, which shows correlations functions for the zero-flow-shear run GKa5 restricted to regions of positive and negative zonal density, as marked in Figure~\ref{fig:ZF},---as is obvious from the tilts of the correlation functions, these regions turn out also to be regions of positive and negative zonal shear (in this context, we should perhaps recall the presence of tilted instantaneous structures in the zero-flow-shear DLM case, noted in Section~\ref{sec:bes_2d_frames}). The overall correlation function in Figure~\ref{fig:gs2_compare_correlation_functions}(d) is an average over several such regions---the opposite tilts average out and the overall radial-reflection symmetry is preserved. 

Since turbulence with mean flow shear will also develop zonal flows, a similar analysis applied to it will show a certain spread in the tilt angle of the correlation function depending on whether the zonal shear locally subtracts from or adds to the mean flow shear. This can matter if the mean flow shear is sufficiently weak.

\subsection{Skewness vs.\ flow shear}
\label{sec:gs2-skew}

\begin{figure}
\centerline{\includegraphics[width=\textwidth]{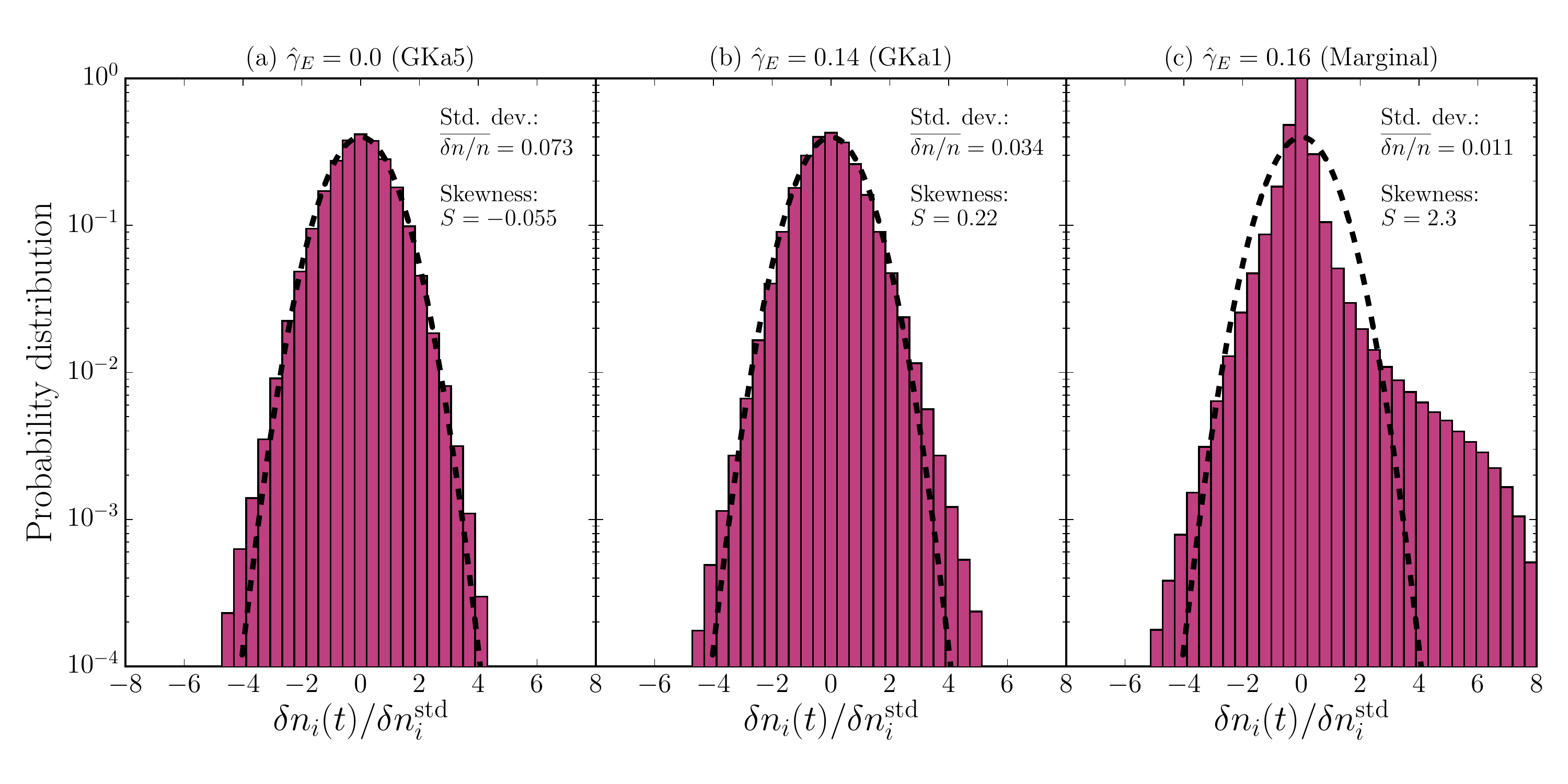}}
\caption{Distribution of the fluctuating density field $\delta n$, normalised by its own standard deviation, for (a) GKa5 (no flow shear), (b) GKa1 (with flow shear, close but not at the nonlinear stability threshold), (c) Marginal (at the threshold) runs (see Table~\ref{tab:flow_shear_scan}). The black-dashed line in each case gives the unit normal distribution. These distributions should be compared with experimental ones with and without flow shear in Figure~\ref{fig:lm_intensity_distribution}. \label{fig:gs2_distributions}}
\end{figure}

The numerical scan in flow shear also reveals that the symmetry of the distribution of the fluctuating density field is broken by the flow shear: the skewness of the distribution increases with the flow shear, as documented in Table~\ref{tab:flow_shear_scan}; the distributions for a run with no flow shear (for which in accordance with the theoretical expectations outlined in Section~\ref{sec:intro}, there is no skew) and runs with two values of the flow shear, one at the nonlinear stability threshold and one just off it, are shown in in Figure~\ref{fig:gs2_distributions}. The skew in the distribution for the Marginal case in Figure~\ref{fig:gs2_distributions}(c) is extremely pronounced, much more so than anything that we have observed experimentally, but the distribution for the case that is only slightly offset from the stability threshold, shown in Figure~\ref{fig:gs2_distributions}(b), is very similar qualitatively to the experimental BLM case in Figure~\ref{fig:lm_intensity_distribution}(a).\footnote{In Section~\ref{sec:LM-intensity-dist}, we considered the distribution of the fluctuating intensity field measured by the BES, whilst here we are considering the distribution of the fluctuating density field generated by the gyrokinetic simulations. There is a certain difference between the two, due to the finite-size PSFs of the BES diagnostic~\cite{Ghim2010}. For a closer comparison with experiment, synthetic BES data can be generated from the gyrokinetic simulations by applying the PSFs of the real diagnostic~\cite{Fox2016}. We will not do this here as we are, in any event, not in a position to make a detailed comparison between simulated and BES-measured turbulent density fields in the same experimental configurations, but are rather looking for qualitative trends. The PSF effects on the distribution functions are considered in Appendix~\ref{sec:sources-of-skew}, where we generate synthetic BES data and show that PSFs do not significantly alter the skewness of the distribution.} Generally speaking, in many runs with various values of $\gEhat$ and $a/L_{T_i}$, we only find distributions with a pronounced tail on the overdensity side in cases that are very close to the threshold (and in all cases that were simulated, we do find asymmetry in favour of over-, rather than underdensities), while the majority of the skewed distributions are similar to that in Figure~\ref{fig:gs2_distributions}(b). This gives us a degree of confidence that the skewness that we have found in the experimentally measured distributions in Section~\ref{sec:LM-intensity-dist} is indeed due to flow shear. It is likely that the fact that this skewness is not very large (and indeed not always measurable) is due to the rather sensitive dependence of it on the distance to the threshold: note the precipitous decline in skewness from the Marginal run to run GKa1, even though the difference in the values of $\gEhat$ for these two cases ($\gEhat=0.16$ and $0.14$, respectively) is small and experimentally would be within measurement errors on the velocity profile (see further discussion in Section~\ref{sec:gs2_asymmetry}). 

\subsubsection{Physics of skewed distributions}
\label{sec:physics-skew} 

Besides encouraging the physical interpretation of the experimental evidence in terms of symmetry breaking associated with flow shear, numerical simulations give us a crucial insight as to how, dynamically, this symmetry breaking occurs. In \cite{vanWyk2016}, for which these simulations were originally carried out, it was shown that the transition to turbulence in these flow-sheared MAST configurations occurs via emergence and accumulation (as equilibrium parameters move deeper into the unstable regime) of long-lived, intense coherent structures, which occupy a small fraction of the volume. This is because turbulence in these equilibria is subcritical---while the system is formally linearly stable, perturbations do grow transiently and, for finite perturbations, this can lead to non-zero nonlinearly saturated turbulent fluctuation levels and heat fluxes. Since finite amplitudes are required for such a state to persist, the only way for the system to have small overall heat fluxes as the threshold is approached is by restricting finite amplitudes to a small fraction of the volume, hence the spatially sparse, but intense structures. Clearly, in terms of the probability distribution of the amplitudes, the presence of such structures and the fact that they give rise to a non-zero overall heat flux, implies skewed distributions. Away from the threshold, the structures become more numerous, overlap, interact, break up and dissolve into a more volume-filling turbulent sea of ``eddies,'' characteristic of standard, unsheared drift-wave turbulence---resulting in approximately symmetric distributions.   

Another important observation afforded us by the numerical simulations but unavailable experimentally (because equilibrium parameters in experiments cannot in general be changed independently) is that, other things being equal, increasing flow shear does have a suppressing effect on the rms amplitude of the turbulent density field: this is documented by the $\overline{\delta n/n}$ values in Table~\ref{tab:flow_shear_scan} (it is not evident in Figure~\ref{fig:gs2_distributions}, because these are distributions of $\delta n$ normalised to its own standard deviation). Thus, as we anticipated in Section~\ref{sec:core_tail}, the tail of the skewed distributions found for the near-threshold, sheared turbulence is, in a sense, a remnant of a wider unsheared/far-from-threshold distribution, containing large-amplitude perturbations that might have existed in a more unstable local equilibrium configuration (although, as argued above, these perturbations have a rather distinct character once only a dwindling number of them are left with a sole responsibility of maintaining turbulent transport; see also Section~\ref{sec:skew-tilt}). 

\subsubsection{Skewness vs.\ tilt: two-component turbulence in simulations} 
\label{sec:skew-tilt}

\begin{figure}
\centerline{\includegraphics[width=\textwidth]{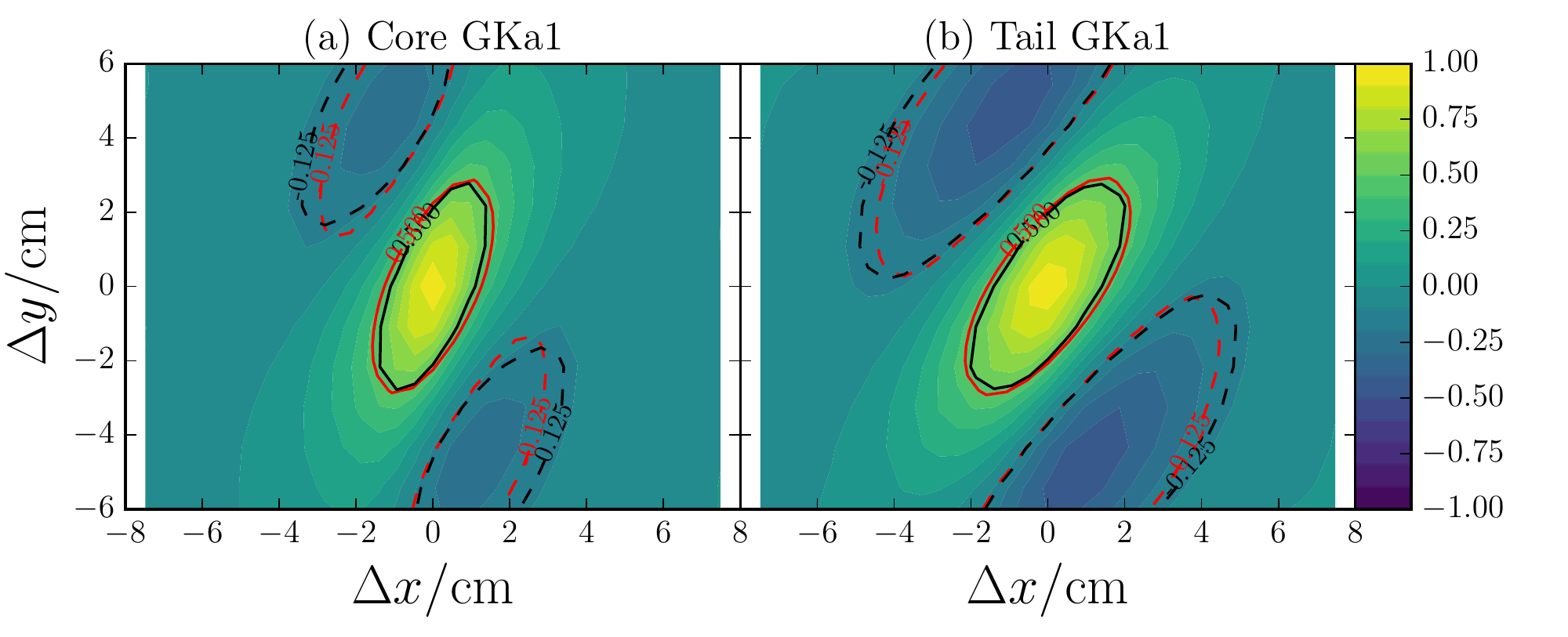}}
\caption{Same correlation function as Figure~\ref{fig:gs2_compare_correlation_functions}(b), for run GKa1, but this time conditioned on (a) lower densities (core of the distribution, $|\delta n| < 2.75$ standard deviations) and (b) higher densities (tail of the distribution, $\max \delta n > 3$ standard deviations) in the way described in Section~\ref{sec:skew-tilt}. The correlation parameters for each case are given in Table~\ref{tab:flow_shear_scan}. These correlation functions can be compared to Figure~\ref{fig:BLM_core_tail} for the experimental BLM case. \label{fig:gs2-conditional}}
\end{figure}

\begin{figure}
\centerline{\includegraphics[width=\textwidth]{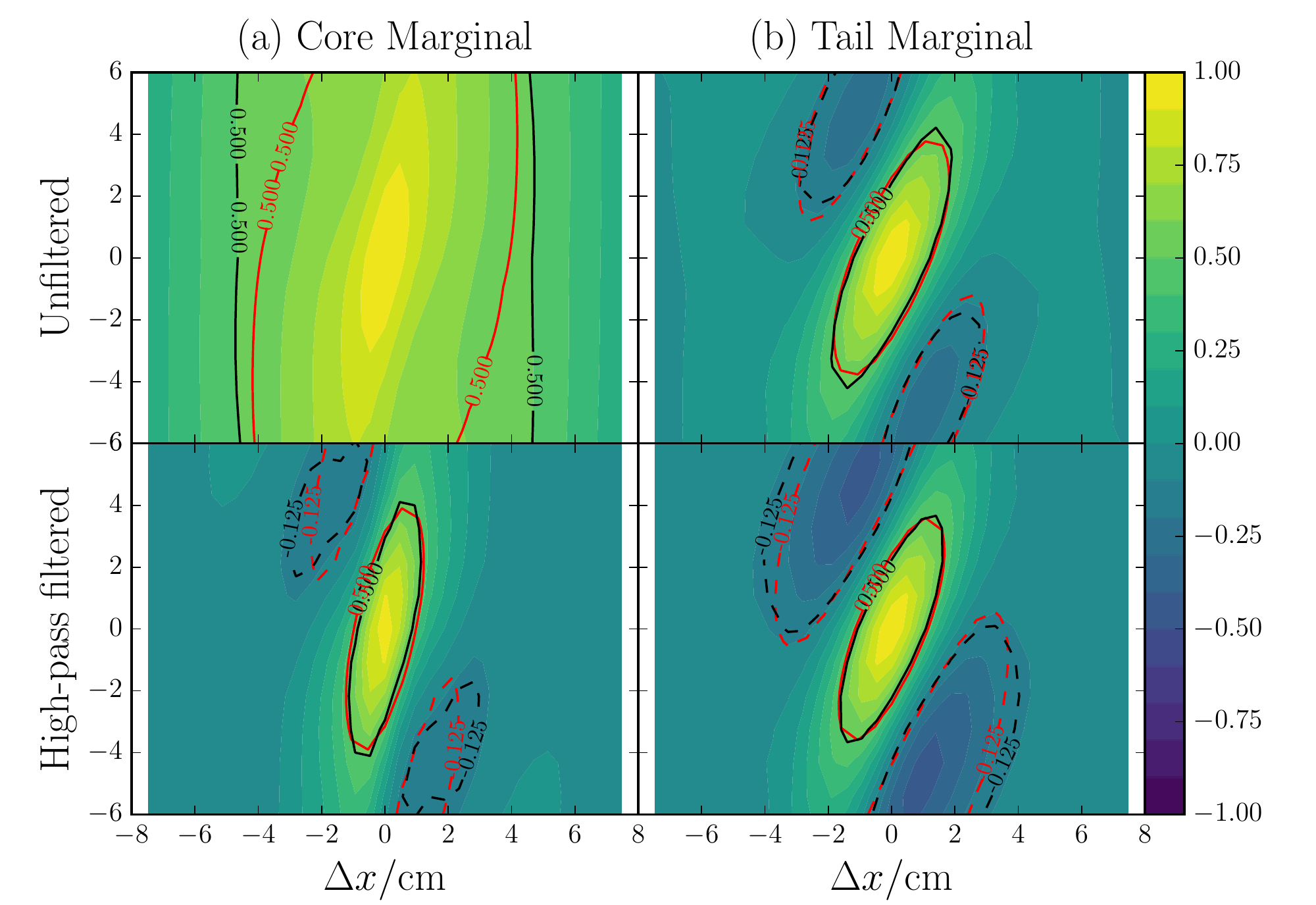}}
\caption{Same as Figure~\ref{fig:gs2-conditional}, but for the Marginal run (see Table~\ref{tab:flow_shear_scan} and Figure~\ref{fig:gs2_compare_correlation_functions}(a)). The bottom row has correlation functions calculated in the standard way from the high-pass-filtered fluctuating density field (as explained at the beginning of Section~\ref{sec:gs2_symmetry_breaking}); the top row has correlation functions of an unfiltered field, i.e., including the zonal density perturbation. The latter manifestly dominates the core population but not the tail.\label{fig:gs2-conditional-marg}}
\end{figure}

In Section~\ref{sec:core_tail}, we conjectured that turbulence in the presence of flow shear is a two-component mix of strong but rare perturbations giving rise to the skew in the distribution and a wider sea of weaker ones, which constitute the core of the distribution and are tilted, more or less passively, by the flow shear. In other words, the tilt and the skew represent symmetry breaking by flow shear occurring in different types of perturbations. 

To test this picture, we introduced conditional correlation functions, designed to measure the spatial correlation parameters associated with the core and the tail of the distribution. Here we perform the same analysis on the numerically simulated fluctuating density fields, with the same condition that frames containing at least one instance of $\delta n$ larger than 3 standard deviations are declared to belong to the tail (for the purposes of this selection, the full artificially large $40\times80$~cm GS2 computational domain is split into $7.48 \times 9.19$~cm subdomains---this is comparable to the $10\times8$~cm BES subarray that we used in Section~\ref{sec:core_tail} and necessary in order for conditioning on the maximum value of $\delta n$ within a subdomain to pick out statistically different patches of turbulence). The results are documented in Table~\ref{tab:flow_shear_scan} and conditional correlation functions for the core and tail of two of the runs with flow shear are shown in Figures~\ref{fig:gs2-conditional} (run GKa1) and~\ref{fig:gs2-conditional-marg} (Marginal run). We see that fluctuations in the tail of the distribution are always less tilted than those in the core. This is similar to the behaviour exhibited by the experimental BLM case in Section~\ref{sec:core_tail}. 

Another qualitative similarity between simulated and real sheared turbulence is that the perturbations in the tail of the distribution, in both cases, have longer radial correlation lengths $\ellx$, not just lower tilt (i.e., smaller $\kx$). Table~\ref{tab:core_tail} shows this for both BLM and IFS cases (even despite the failure of the latter to follow the trend as far as its tilt angle is concerned) and Table~\ref{tab:flow_shear_scan} confirms the same trend for our numerical runs with varying flow shear; it is also quite obvious from the plots of the conditional correlation functions in Figures~\ref{fig:gs2-conditional} and~\ref{fig:gs2-conditional-marg}. 

Comparing these two figures, we note also a sizeable increase of the poloidal correlation length in the core of the distribution as the threshold is approached (see Table~\ref{tab:flow_shear_scan}). Interestingly, in the Marginal run, the core fluctuations are so weak, that the zonal density perturbation (in our definition of the fluctuating density field, it is removed by time averaging/high-pass frequency filtering; see Section~\ref{sec:ZF}) becomes comparable to them---so much so that, as we see in Figure~\ref{fig:gs2-conditional-marg}, it dominates the conditional core correlation function (although not the total one or the tail) if the high-pass frequency filter is removed (suggesting what is perhaps a promising experimental direction of enquiry).

\subsection{Restoration of symmetry far from threshold}
\label{sec:gs2_asymmetry}

In our discussion of both experimental and numerical results so far, we have repeatedly anticipated the argument that what truly matters for the degree to which the symmetry of the turbulence is broken is how far from the (nonlinear) stability threshold the system is. Thus, changing the flow shear in Sections~\ref{sec:gs2_symmetry_breaking} and \ref{sec:gs2-skew} led to significant changes in behaviour because this took the system closer to or farther from the marginal case. 

\begin{table}
\centering 
\begin{tabular}{r|rrrrrrrr} 
Run name &Marginal&GKb1&GKb2&GKb3&GKb4&GKb5&GKb6&GKb7\\ 
$a/L_{T_i}$& $4.8$& $4.9$& $5.0$& $5.1$& $5.2$& $6.0$& $7.0$& $8.0$\\ 
\hline\rule{0pt}{1.2em}
$Q_i/Q_\mathrm{gB}$& $1.4$& $5.2$& $7.9$& $9.8$& $14.8$& $44.0$& $92.3$& $142.2$\\  
$\delta n/n$& $0.011$& $0.021$& $0.027$& $0.030$& $0.035$& $0.069$& $0.11$& $0.14$\\ 
Skewness& $2.3$& $0.50$& $0.40$& $0.28$& $0.24$& $0.075$& $0.094$& $0.12$\\ 
\hline\rule{0pt}{1.2em}
$\ellx/\rho_i$& $3.5$& $3.1$& $3.2$& $3.2$& $3.3$& $3.7$& $3.6$& $3.6$\\ 
$\elly/\rho_i$& $10$& $10$& $9.9$& $9.4$& $9.3$& $8.6$& $9.3$& $9.3$\\ 
$\kx\rho_i$& $-0.51$& $-0.44$& $-0.43$& $-0.42$& $-0.41$& $-0.34$& $-0.24$& $-0.24$\\ 
$\ky \rho_i$& $0.24$& $0.25$& $0.25$& $0.26$& $0.27$& $0.29$& $0.27$& $0.26$\\ 
Tilt $\Theta/\mathrm{deg}$& $69$& $65$& $64$& $63$& $62$& $54$& $47$& $48$\\ 
\hline\rule{0pt}{1.2em}
$\tauc$& $16$& $13$& $13$& $12$& $12$& $8.7$& $6.7$& $7.0$\\  
\hline\rule{0pt}{1.2em}
CORE &&&&&&&&\\
$\ellx/\rho_i$& $2.7$& $2.6$& $2.7$& $2.7$& $2.8$& $3.2$& $3.1$& $3.2$\\ 
$\elly/\rho_i$& $9.3$& $8.8$& $8.6$& $8.2$& $7.9$& $7.4$& $7.5$& $7.4$\\ 
$\kx\rho_i$& $-0.56$& $-0.49$& $-0.48$& $-0.47$& $-0.46$& $-0.38$& $-0.29$& $-0.31$\\ 
$\ky\rho_i$& $0.18$& $0.25$& $0.25$& $0.26$& $0.27$& $0.29$& $0.27$& $0.26$\\ 
Tilt $\Theta/\mathrm{deg}$& $75$& $67$& $66$& $66$& $64$& $53$& $52$& $55$\\ \hline 
\hline\rule{0pt}{1.2em}
TAIL &&&&&&&&\\
$\ellx/\rho_i$& $3.6$& $3.7$& $3.9$& $4.0$& $3.9$& $4.4$& $4.2$& $4.1$\\ 
$\elly/\rho_i$& $10$& $10$& $9.8$& $9.8$& $9.9$& $8.9$& $9.7$& $9.9$\\ 
$\kx\rho_i$& $-0.50$& $-0.42$& $-0.40$& $-0.38$& $-0.38$& $-0.30$& $-0.19$& $-0.17$\\ 
$\ky\rho_i$& $0.26$& $0.27$& $0.26$& $0.27$& $0.28$& $0.30$& $0.28$& $0.28$\\ 
Tilt $\Theta/\mathrm{deg}$& $67$& $62$& $62$& $60$& $58$& $51$& $40$& $36$\\ 
\end{tabular} 
\caption{Same as Table~\ref{tab:flow_shear_scan}, but for $\gEhat=0.16$ and a sequence of values of the ion-temperature 
gradient $a/L_{T_i}$. \label{tab:rlti_scan}}
\end{table}

This notion is easily confirmed if we fix the flow shear at what in the previous sections was the marginal value, $\gEhat=0.16$, and change instead the ion-temperature gradient (ITG) $a/L_{T_i}$. A sequence of runs within such a parameter scan is documented in Table~\ref{tab:rlti_scan}. The trend that emerges is very clear: as the ITG is increased from the marginal value ($a/L_{T_i}=4.8$), both the tilt angle of the correlation function and the skewness of the distribution of the fluctuating densities diminish (the decline in skewness is particularly strong). Thus, the symmetry is broken less strongly.\footnote{Technically, the symmetry is either broken or it is not. However, we resort to the colloquial phrasing referring to the `strength' of the symmetry breaking as a synonym for the size of the characteristics (tilt and skew) that indicate the symmetry has been broken.} The situation in this sense is rather similar to what happened in the $\gEhat$ scan (Table~\ref{tab:flow_shear_scan}) with decreasing flow shear. One might argue that, as turbulence becomes stronger away from the threshold (with the distance from the threshold measured either in $\gEhat$ or in $a/L_{T_i}$), the flow shear ceases to be an important factor and the symmetry formally broken by it is effectively restored. 

\subsubsection{Tilt, flow shear and life time}
\label{sec:gs2_away_from_marginal}

It is worth discussing a little further how the effective restoration of symmetry happens for the tilt of the correlation functions. We argued at the end of Section~\ref{sec:intro-tilt} and again in Section~\ref{sec:tilt-taulife} that the flow shear did not need to be dynamically important in order to induce a tilt in the correlation function. In such a ``passive'' tilting scenario, the tilt $\tan\Theta = \kx/\ky$ is simply proportional to $\gE$, with the constant of proportionality being the life time $\tauc$ of the perturbations, as per (\ref{low-shear}) or (\ref{tauc-def}). This life time, defined by (\ref{tauc-def}) and given in Table~\ref{tab:rlti_scan}, becomes shorter as the ITG is increased. This is also manifest in Figures~\ref{fig:taulife_vs_gE}(a) and (b)---the latter figure is a scatter plot of $\tauc$ for a large set of values of $\gEhat$ and $a/L_{T_i}$ against the gyro-Bohm-normalised ion heat flux $\Qgb$, which is a good measure of distance to threshold (it is zero at the threshold and increases steeply away from it \cite{vanWyk2016}). The reason correlation times shorten is that larger ITG implies stronger drive and so higher turbulent amplitudes: ``eddies'' turn over faster.\footnote{Very far from the threshold, in ITG turbulence without flow shear (which, far from the threshold, presumably does not matter anyway), the correlation times become independent of the ITG because turbulence is controlled by the ``critical balance'' between linear, nonlinear and streaming time scales \cite{Barnes2011}, but here we are discussing cases that are still relatively close to the threshold (although one might argue that $a/L_{T_i}=7$ and $8$ in Figure~\ref{fig:taulife_vs_gE}(a) already have fairly similar values of $\tauc$).} Shorter $\tauc$ then implies smaller tilt angle ($\tan \Theta \sim \gE\tauc$) farther from marginality. 

In physical terms, the flow shear ceases to be important if it is smaller than the local, fluctuating velocity shear acting on a turbulent perturbation and caused by its fellow perturbations. This nonlinear fluctuating shear must, in saturation, be comparable to some appropriate linear rate at which the perturbations are replenished by the ITG drive. In this sense, the competition between the life time of the turbulence and the flow shear as the threshold is approached is reminiscent of the so-called Waltz rule~\cite{Waltz1994}: that in order to affect (suppress) the turbulence, the flow shear must be similar to, or greater than, the linear growth rate (we recall, however, that the turbulence that we are dealing with here is subcritical \cite{vanWyk2016} and so formally there is no linear growth rate; what effective quantity from the linear theory should be used instead is a matter of current research interest \cite{Schekochihin2012,Highcock2012,vanWyk2016b}). 

\subsubsection{Tilt and skewness vs.\ distance from threshold}
\label{sec:restore_sym}

\begin{figure}
\centering
\begin{tabular}{cc}
\includegraphics[width=0.45\textwidth]{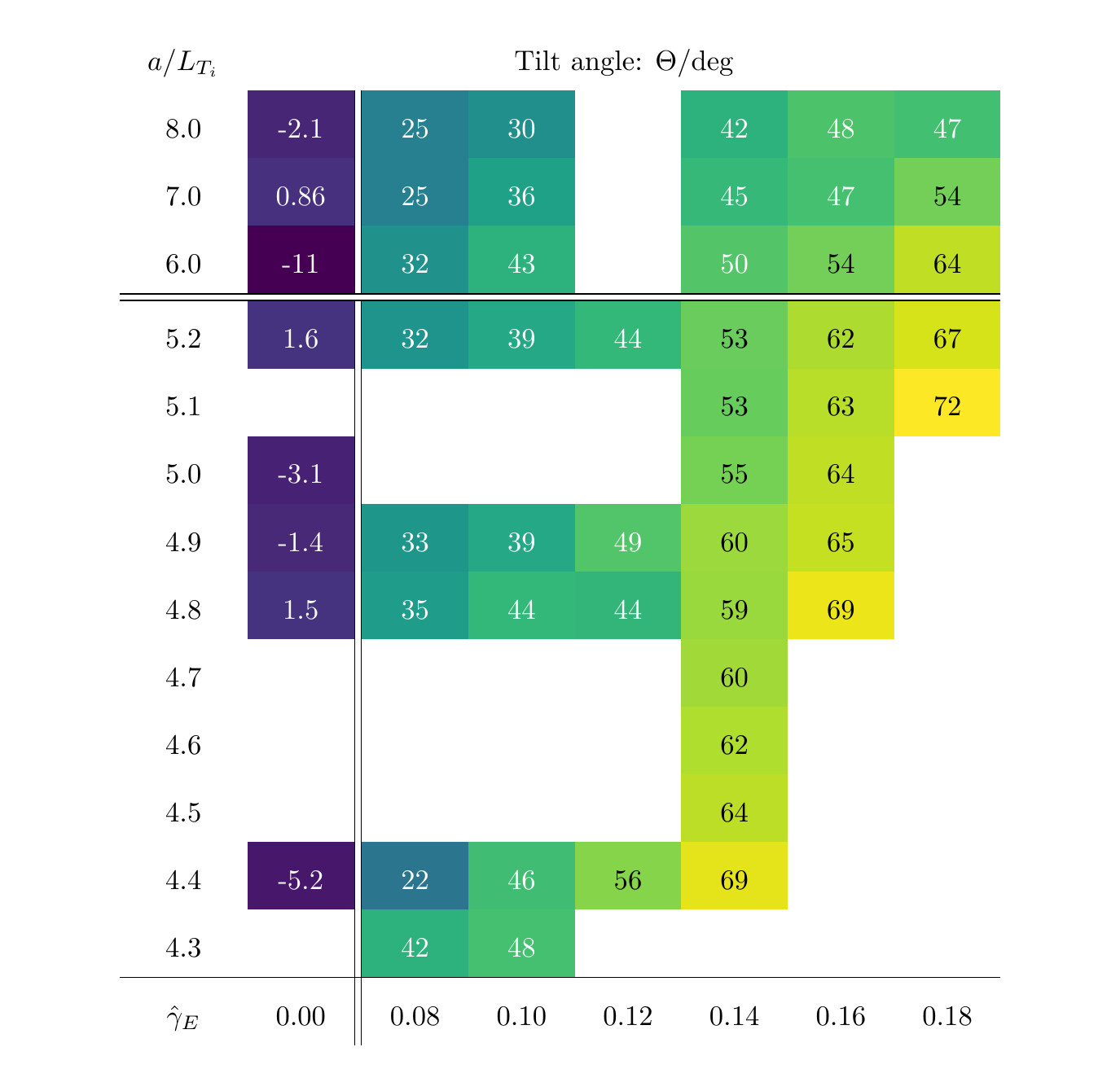} & 
\includegraphics[width=0.45\textwidth]{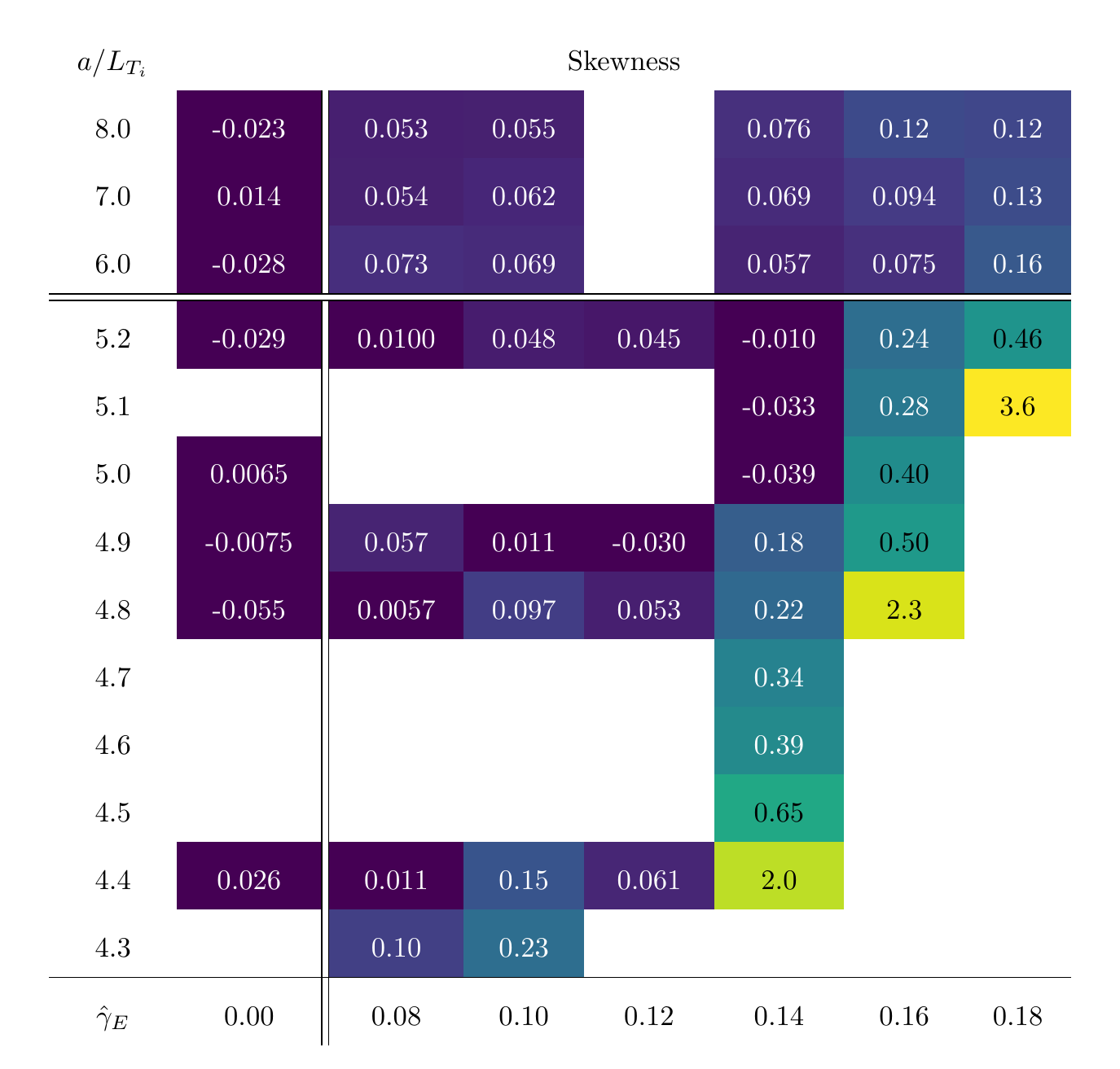}\\
(a) Tilt vs.\ $\gEhat$ and $a/L_{T_i}$ & (b) Skewness vs.\ $\gEhat$ and $a/L_{T_i}$\\\\
\includegraphics[width=0.45\textwidth]{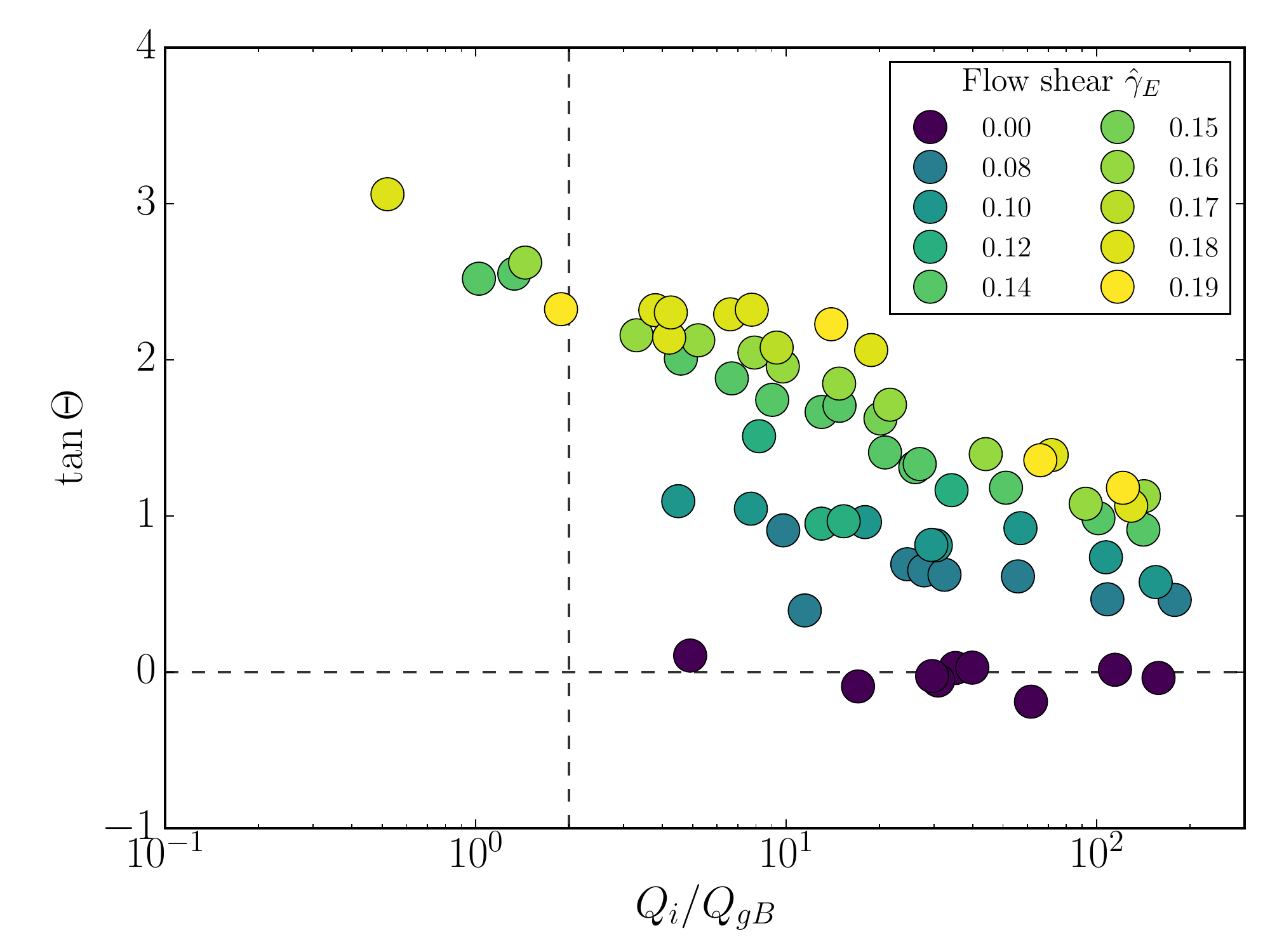} & 
\includegraphics[width=0.45\textwidth]{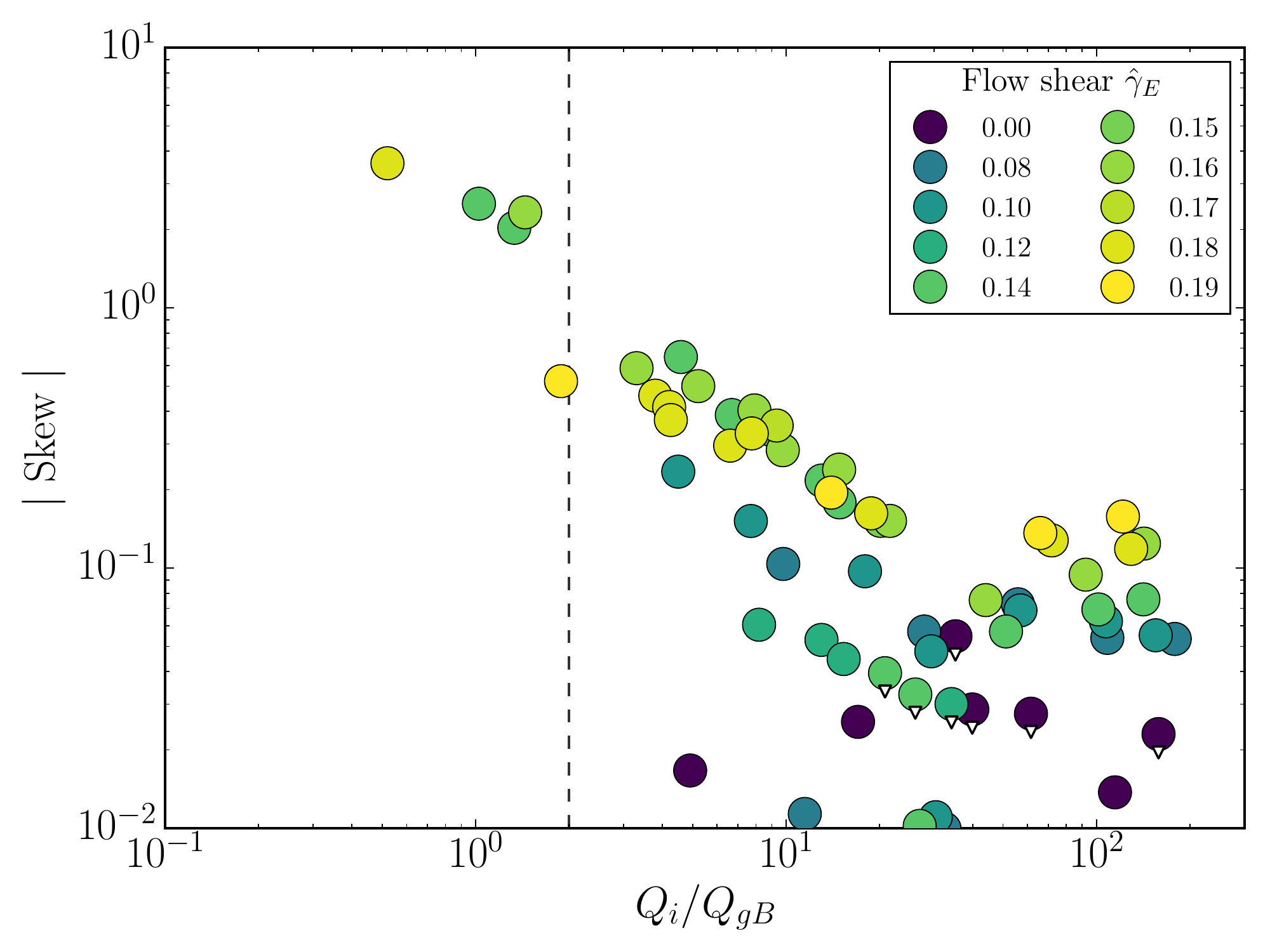}\\
(c) Tilt vs.\ $\Qgb$ & (d) Skewness vs.\ $\Qgb$\\
\end{tabular}
\caption{(a) Tilt angle of the correlation function and (b) skewness of the distribution of the fluctuating density field vs.\ flow shear $\gEhat$ and ITG $a/L_{T_i}$ in a range of values of these parameters around the nonlinear stability threshold. The tilt and the skewness are replotted in (c) and (d), respectively, vs.\ the gyro-Bohm-normalised ion heat flux, with data points coloured according to the value of $\gEhat$. These plots contain data for $\gEhat\in[0,0.19]$ and ITG $a/L_{T_i}\in[4.3,8]$, covering the entire database of the runs carried out in \cite{vanWyk2016}. In (d), inverted triangles mark the cases where skewness is negative.\label{fig:tilt-skew}}
\end{figure}

Figures~\ref{fig:tilt-skew}(a) and (b) show the tilt angle of the correlation function and skewness of the distribution of the fluctuating density field for a range of values of the flow shear $\gEhat$ and ITG $a/L_{T_i}$ around the nonlinear stability threshold and departing from it---this extends the $\gEhat$ and $a/L_{T_i}$ scans documented in Tables~\ref{tab:flow_shear_scan} and~\ref{tab:rlti_scan}. The conclusion remains the same: away from the threshold, symmetry is effectively restored---particularly quickly for the distribution functions (as we argued in Section~\ref{sec:physics-skew}, the restoration of symmetry of the distribution of fluctuating densities far from the threshold occurs because intense coherent structures that dominate the transition to turbulence and are responsible for the skewed distributions become too numerous for individual survival, interact and break each other apart \cite{vanWyk2016}). 

As we already did in Section~\ref{sec:gs2_away_from_marginal} for the life time of the perturbations, we quantify the dependence of the measures of asymmetry (tilt and skewness) on the distance to threshold in terms of their dependence on the gyro-Bohm-normalised heat flux $\Qgb$, which vanishes at the threshold but increases strongly and monotonically away from it \cite{vanWyk2016} and thus constitutes a good order parameter. Scatter plots of the tilt and the skewness vs.\ $\Qgb$ are shown in Figures~\ref{fig:tilt-skew}(c) and (d), respectively, for the entire data base of numerical simulations originally carried out in \cite{vanWyk2016}. We see that, modulo some scatter, both the tilt and the skewness are approximately functions of the distance to the threshold (to be precise, Figure~\ref{fig:taulife_vs_gE}(b) suggests that it is the life time $\tauc = \tan\Theta/\gE$, rather than the tilt, that is a nearly single-valued function of this distance, at least when measured sufficiently far from the threshold). The case of $\gEhat=0$ is the exception to this rule, as expected theoretically, being subject to the exact symmetry (\ref{symmetry-transforms}): namely, at zero flow shear, both the tilt and the skew are essentially zero regardless of how marginal, or otherwise, the system is.  

\section{Discussion}\label{sec:conclusions}

To summarise, we started with a premise that, as the presence of the flow shear appears to be quite strongly correlated with larger temperature gradients both in MAST \cite{Ghim2014} and in numerical simulations \cite{Highcock2012,vanWyk2016}, the effect of the shear on the local structure of plasma turbulence must be detectable. Theoretically, we argued (in Section~\ref{sec:intro}) that flow shear would break the reflection symmetry (\ref{symmetry-transforms}) \cite{Parra2011,Sugama2011} and that statistically this symmetry breaking would be actualised in the form of skewed distributions (Section~\ref{sec:intro-skew}) and tilted correlation functions (Section~\ref{sec:intro-tilt}) of the fluctuating density field. We did then find both of these signatures of symmetry breaking experimentally (Sections~\ref{sec:sb-skew} and~\ref{sec:sb-tilt}, respectively), at least as a proof of principle---in specific examples of turbulent density fields in MAST with and without flow shear. 

For a number of reasons to do with spatial resolution of the MAST BES system and with the range of equilibrium parameters available, we cannot, at this point, have a broad parameter-scan study and so we sought reassurance, validation and further physical insight from a set of numerical simulations of MAST turbulence performed by \cite{vanWyk2016} and covering a range of flow shears and ion-temperature gradients in a MAST-relevant equilibrium configuration. Again we found both skewed fluctuating-density distributions (Section~\ref{sec:gs2-skew}) and tilted correlation functions (Section~\ref{sec:gs2_symmetry_breaking}), qualitatively similar to the experimental ones. 

With the aid of the simulations, we were able to establish that, generally speaking, the symmetry breaking was most pronounced in the vicinity of the (nonlinear) stability threshold, whereas away from it, symmetry was gradually restored (Section~\ref{sec:restore_sym})---essentially because flow shear became dynamically less and less relevant far from the threshold (Section~\ref{sec:gs2_away_from_marginal}). The dependence of the symmetry breaking on the distance from threshold---especially of the skewness of the fluctuating-density distributions---turned out to be quite strong. Experimentally, while we expect to find turbulent states in the general vicinity of the stability boundary (as indeed confirmed in \cite{vanWyk2016}), it is not reasonable to expect them to be exactly on it. The degree to which symmetry is found to be broken can then be viewed as an indicator of how close to the threshold any given measured instance of turbulence is. 

We have also argued that tilted correlation functions provide a versatile diagnostic that could be used to probe the effective life time (correlation time) of turbulent structures (Sections~\ref{sec:corr_times}, \ref{sec:tilt-taulife}, and~\ref{sec:gs2_away_from_marginal}) and the local zonal shear (Section~\ref{sec:ZF}). Furthermore, when conditioned on the amplitude of the fluctuating density, these correlation functions appeared to reveal, both experimentally (Section~\ref{sec:core_tail}) and numerically (Section~\ref{sec:skew-tilt}), that sheared, skewed, near-threshold turbulence could be thought of as consisting of two components: longer-lived, spatially larger, more intense, but statistically rarer structures and a general sea of smaller, weaker, but statistically volume-filling fluctuations. Indeed, that spatially sparse coherent structures dominate the transition to turbulence in the physical situation considered here was shown by \cite{vanWyk2016} (see the summary of their argument in Section~\ref{sec:physics-skew})---what we have found here is that the statistical signature of these structures is the breaking of symmetry (skewness) of the distribution function of the fluctuating density, or, conversely, that the physical way in which the system takes advantage of the breaking of symmetry is the emergence of these structures. The tilting of the correlation functions is a better understood and, arguably, more straightforward phenomenon (Section~\ref{sec:intro-tilt}) affecting both strong and weak perturbations, the latter more than the former (see Sections~\ref{sec:core_tail} and \ref{sec:skew-tilt}). 

It is quite clear that, experimentally, a much more extensive study is called for. The trends revealed by the numerical simulations serve as a guideline to the range of questions that can be asked and of parameter dependences that could be verified or falsified. It would be interesting to learn in what circumstances fluctuating-density distributions are skewed towards underdensities, rather than overdensities (exemplified by one of our experimental cases; see Section~\ref{sec:sb-skew}). The nature and role of zonal flows generally and in subcritical turbulence with flow shear in particular (such as the MAST turbulence that we have investigated), remain a fascinating subject, which perhaps could be tackled experimentally using some of the tools developed here (see Section~\ref{sec:ZF}) or other methods \cite{FieldIAEA2014}. More generally, a full experimental (as well as numerical and theoretical) understanding of the transition to turbulence and of the resulting turbulent states in subcritically unstable, sheared tokamak turbulence is a worthwhile goal, important for practical purposes of optimising tokamaks and perhaps learning how to manipulate stability boundaries, but also appealing to the more curiosity-driven angels of our nature.   

\section*{Acknowledgements}
This work was carried out within the framework of the EUROfusion Consortium. It received funding from the Euratom research and training programme 2014-18 under grant agreement No 633053 and from the RCUK Energy Programme [grant number EP/I501045]. The views and opinions expressed herein do not necessarily reflect those of the European Commission. Financial support was also provided to MFJF by Merton College, Oxford. YcG was supported by the National R\&D Program through the National Research Foundation of Korea (NRF) funded by the Ministry of Science, ICT and Future Planning (Grant No.\ 2014M1A7A1A01029835) and by the KUSTAR-KAIST Institute, KAIST, Korea. The work of AAS was supported in part by grants from UK STFC and EPSRC. We thank M.~Barnes, S.~Cowley, W.~Dorland, E.~Highcock and C.~Roach for many inspiring discussions of plasma turbulence and its modelling. 

\bibliographystyle{phaip}
\bibliography{references_thesis.bib}

\appendix

\section{Measuring skewed distributions}
\label{sec:sources-of-skew}

Experimental measurement of skewed (or otherwise) distributions of the fluctuating intensity fields is complicated by a number of extraneous effects, which we detail and describe how to correct for in Appendix~\ref{sec:correct_skew}. In Appendix~\ref{sec:app_PSF}, we extend the work of~\cite{Fox2016} to determine the effect of PSFs on the skewness of the fluctuating-field distributions. 

\begin{figure}
	\centering
	\includegraphics[width=\textwidth]{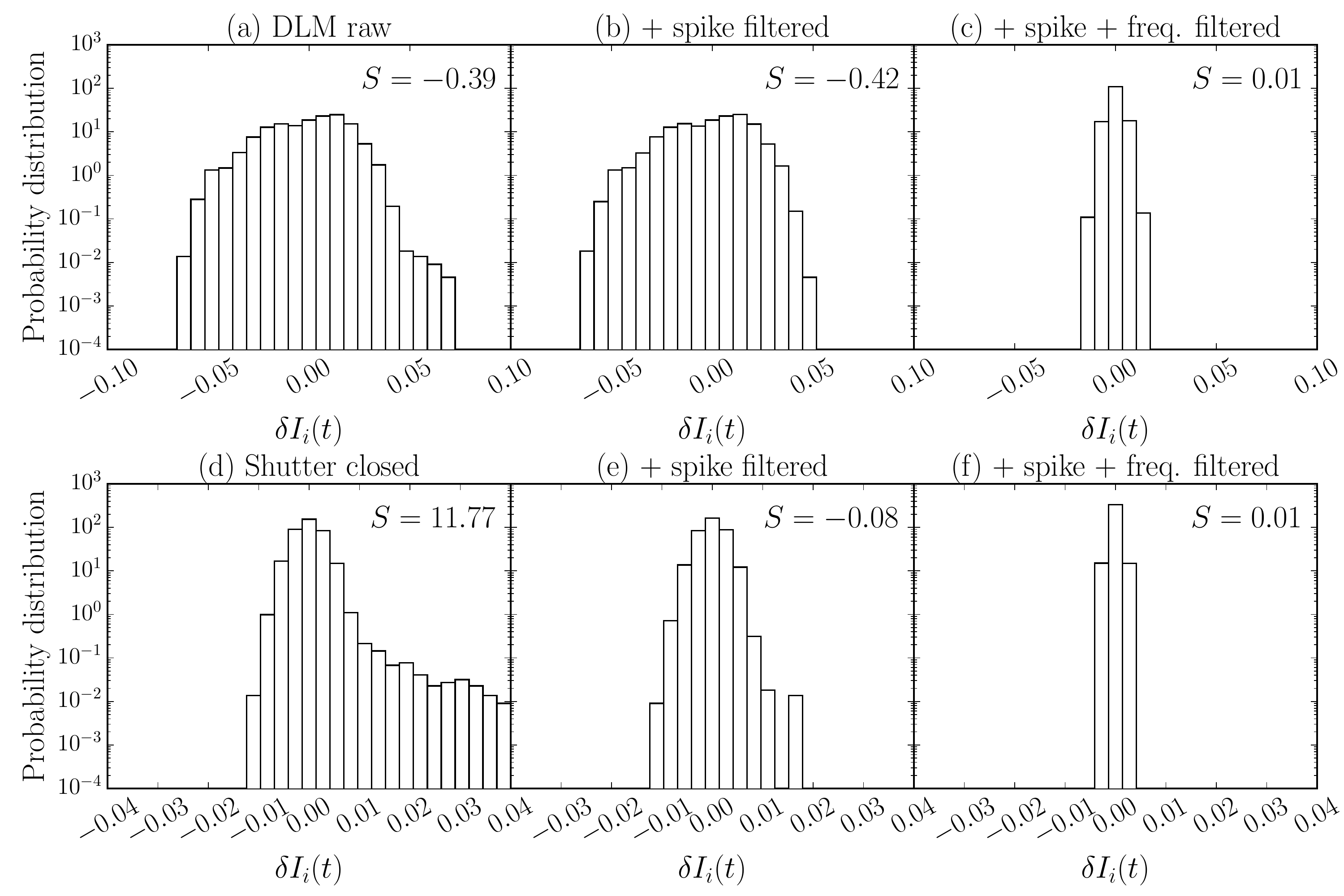}
	\caption{Probability distribution of the intensity perturbation $\delta I_i$ (unnormalised) for (a) the DLM case (see Section~\ref{sec:experimental-configuration}) for $2\times4$ BES channels, as in in Figure~\ref{fig:lm_intensity_distribution}(b); (b) the same as (a) but spike-filtered; (c) the same as (a) but spike- and then bandpass-filtered in the range $20-100~\mathrm{kHz}$; (d) the detected signal of the BES diagnostic when the shutter is closed, measured during shot \#27367 at $t\in[250,252]~\mathrm{ms}$, using the $5\times4$ channels inner BES subarray; (e) the same as (d) but spike-filtered; (f) same as (d) but spike- and then bandpass-filtered in the range $20-100~\mathrm{kHz}$. Note the different $x$-axis scales in (a-c) and (d-f).\label{fig:background_check}}
\end{figure}

\subsection{Accounting and correcting for spurious skewness}
\label{sec:correct_skew}
The distribution of fluctuating intensities of the raw BES signal can be skewed for two main reasons not associated with the symmetry breaking of the turbulent field: the presence of high-energy radiation and the slow variation of the intensity signal. We discuss these effects by considering Figure~\ref{fig:background_check}(a), where the distribution of (unnormalised) fluctuating intensities from the raw BES signal for the DLM case (see Section~\ref{sec:LM-intensity-dist}) is plotted. 

First, let us consider the smaller (in this case) of the two effects: the low-probability tail on the right-hand side of the distribution in Figure~\ref{fig:background_check}(a). This tail is caused by high-energy radiation (neutron, gamma ray, or hard $X$-ray) impinging on the avalanche-photodiode detector (APD) of the BES camera and producing spikes in the time series that are isolated to a single detector channel. 

We can observe this effect more clearly by analysing BES data from another MAST shot (\#27367), where the shutter over the optical system of the BES diagnostic was closed, and so no beam-emission photons were able to reach the APD. The distribution of unnormalised fluctuating intensities for this ``shutter-closed'' case is plotted in Figure~\ref{fig:background_check}(d). It can be seen to be made up of two components: the first is a Gaussian distribution of photon (electronic) noise, the second is a tail of positive $\delta I$ due to the high-energy radiation, causing large positive skewness $S=11.77$. 

The sum of two randomly distributed independent variables (here, beam-emission photons and high-energy radiation) has a probability distribution that is the convolution of the probability distributions of the two variables (see, e.g., \cite{BendatPiersol}). Therefore, formally, knowing (having measured) the probability distribution of the radiation spikes, it is possible to deconvolve this probability distribution from an experimentally measured probability distribution. However, this is not possible in practice because the actual number of large-amplitude radiation spikes is small (typically at most one or two per channel in a $2~\mathrm{ms}$ time window). As a result, we resort to a cruder method, but an effective one, of removing the radiation spikes from the time series by identifying large (above a certain threshold) differences between the intensity at one point in time and the next, and then replacing this large-intensity value with the value of the neighbouring point\footnote{The radiation spikes can persist over a number of sampling times (each equal to $0.5~\mu\mathrm{s}$), so the spike-filtering algorithm is designed to fill in up to 4 time points with neighbouring values.}~\cite{Field2012}. The result of this ``spike filtering'' can be seen in Figure~\ref{fig:background_check}(e), where the large tail of the distribution in Figure~\ref{fig:background_check}(d) has been successfully removed by this method.

However, our spike filter is not able to remove all of the radiation spikes, especially ones with smaller amplitudes. These remaining low-amplitude radiation spikes cause high-frequency variations in the time series. Therefore, a low-pass filter can be used to remove them. A low-pass filter is also necessary to remove high-frequency (above $100~\mathrm{kHz}$) noise from the signal (see Figure 11 of~\cite{Field2012}). In order to account for the slow variation of the intensity signal (which will be discussed shortly), a high-pass filter is also required. Therefore, we use a bandpass filter in the range $20-100~\mathrm{kHz}$ on the BES-measured time series. The outcome is shown in Figure~\ref{fig:background_check}(f). The skewness of the fluctuating-intensity distribution in this figure is nearly zero, demonstrating that this process can successfully account for the skewness caused by high-energy radiation and gives us an estimate of the uncertainty with which we can measure the skewness of the distribution of the turbulent field (in the absence of other effects; see Appendix~\ref{sec:app_PSF}). Returning to the DLM case, in Figure~\ref{fig:background_check}(b), we see that the spike filter successfully removes the low-probability tail. 

Now we consider the second spurious source of skewness in the distribution: the slow ($\sim\mathrm{ms}$) time evolution of the intensity signal due to changes in the equilibrium and slow MHD modes ($>50~\mu\mathrm{s}$). If, during a given time interval, the (running) mean intensity signal varies unevenly (for example, decreases slightly for the last third of the time interval), then the mean over the entire time interval $\langle I_i(t)\rangle$ will be weighted by this variation. The distribution of the fluctuating part of the intensity signal, $\delta I_i(t) = I_i(t) - \langle I_i(t)\rangle$, will then be skewed. Therefore, a high-pass filter is used to remove all variations of the intensity signal with frequencies below $20~\mathrm{kHz}$. We demonstrate the effect of using such a filter (in combination with the low-pass filter discussed above) on the DLM case, which has, on inspection of the raw time series, a slow, uneven variation in the intensity signal over a $2~\mathrm{ms}$ time interval. In Figure~\ref{fig:background_check}(c), the $20-100~\mathrm{kHz}$ bandpass filter reduces the skewness in the distribution significantly, compared to that in Figure~\ref{fig:background_check}(b). We have now recovered the distribution that was shown in Figure~\ref{fig:lm_intensity_distribution}(b).

We finish by noting another possible source of spurious skewness in the distribution of fluctuating intensities, the background emission from the plasma. Intensity distributions for the cases either when there is no neutral beam heating or when only the SW neutral beam is active (the BES images the SS beam), are similar to that of the shutter-closed case in Figure~\ref{fig:background_check}(d). Therefore, we do not expect the background plasma emission to produce a skewness larger than that caused by the high-energy radiation, and that we have been able to account and correct for.

\begin{figure}
	\centering
	\includegraphics[width=\textwidth]{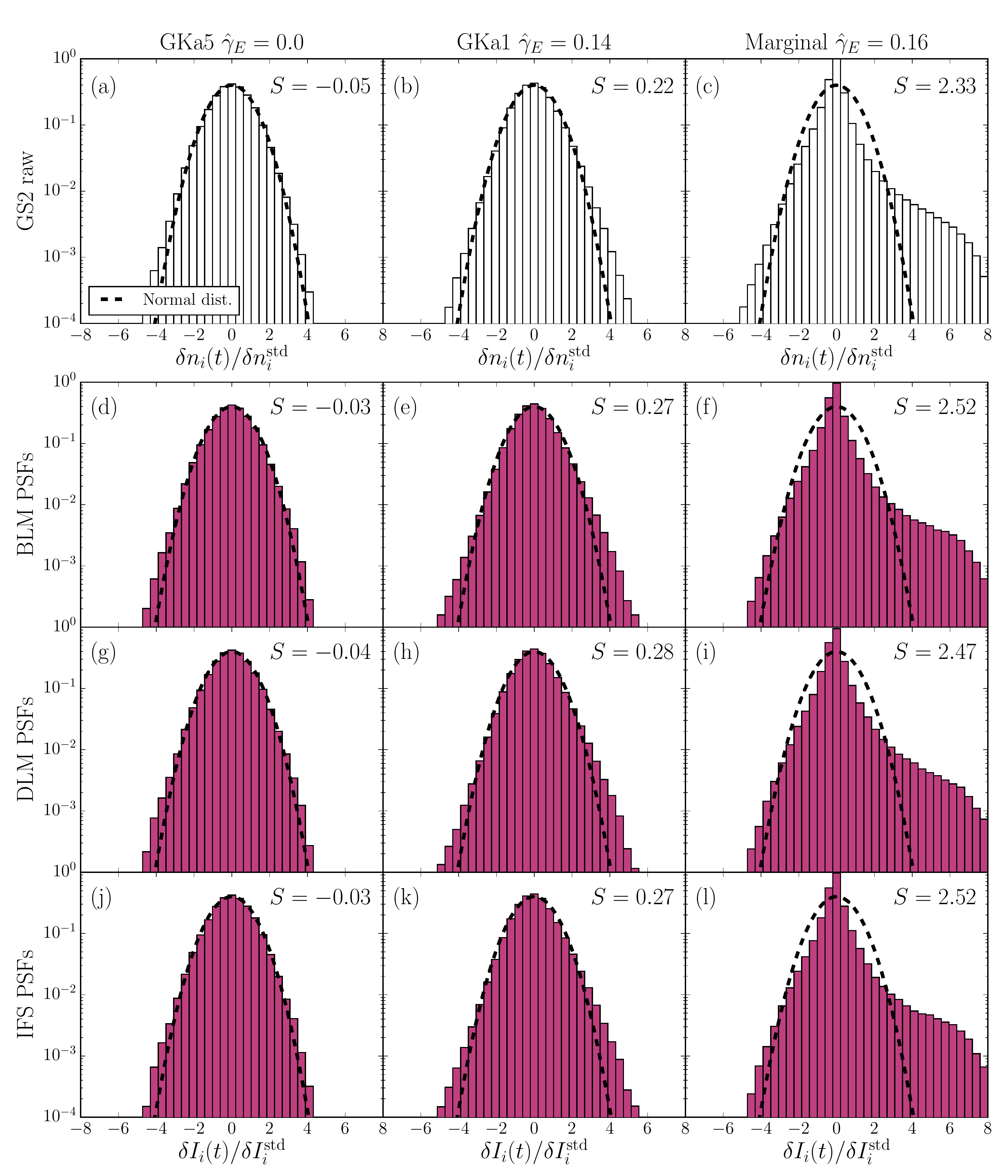}
	\caption{Effect of PSFs on the distribution of amplitudes from the GKa5, GKa1, and Marginal simulations. The intensity fields (d-l) are calculated from the density fields (a-c) by evaluating (\ref{PSF_effects}). The dashed line in each plot is the normal distribution. \label{fig:psf_distribution}}
\end{figure}

\subsection{PSF effects}
\label{sec:app_PSF}
We now consider PSF effects on the distribution of the fluctuating field. In the first row of Figure~\ref{fig:psf_distribution}, panels (a-c), we reproduce the three distributions of the numerically simulated fluctuating density field already plotted in Figure~\ref{fig:gs2_distributions} (these are the distributions for the GKa5, GKa1, and Marginal runs). In each of the following rows, we plot the distributions of the fluctuating intensity field calculated from the density field used in panels (a-c) and using the PSFs of the BES diagnostic calculated for each experimental case that we considered in Section~\ref{sec:LM-intensity-dist}: the second row, panels (d-f), is for the BLM case, the third row, panels (g-i), is for the DLM case, and the fourth row, panels (j-l), is for the IFS case. The relationship between the density field, the intensity field and the PSFs is given by
\begin{equation}
\label{PSF_effects}
\delta I_i = \int P_i(r-r_i,Z-Z_i) \beta \delta n(r,Z) \mathrm{d}r \mathrm{d}Z,
\end{equation}
where $P_i(r-r_i,Z-Z_i)$ is the PSF of channel $i$, calculated according to~\cite{Ghim2010}, and $\beta$ is related to the atomic physics of the line emission and is a weak function of density~\cite{Hutchinson2002} (for our three experimental cases, it is approximately constant: $\beta=0.7$).

For the run with zero flow shear (GKa5), we see from the first column of Figure~\ref{fig:psf_distribution} that the skewness is changed from $-0.05$ without PSFs to at most $-0.03$ when applying the PSFs. In the second column of Figure~\ref{fig:psf_distribution}, we see similar behaviour for the runs with a moderate flow shear (GKa1), with an increase of skew of $0.06$. For the Marginal run, the skewness is again increased by the PSFs, but by an even lesser amount, in the 10\% range (irrelevant compared to the true skewness for these runs). It is clear that in the runs with non-zero flow shear the change in skewness due to the PSFs is small compared to the true value of skewness. These results suggest that the difference in skewness between the three experimental cases considered in Section~\ref{sec:LM-intensity-dist} cannot be due solely to PSF effects. In fact, we see that all three of the sets of PSFs taken from the experimental cases have similar effects on the distribution of the fluctuating field. 

\end{document}